\documentclass[12pt,a4paper]{article}
\usepackage[dvipdfmx]{graphicx}
\usepackage{amsmath,amssymb}
\usepackage{authblk}
\usepackage{comment}
\usepackage{here}
\usepackage{docmute}
\usepackage{simplewick}
\usepackage{cite}
\usepackage{layout}
\usepackage[margin=2.4cm]{geometry}
\usepackage{setspace} 
\usepackage{color}

\usepackage{ulem}
\usepackage{cancel}

\title{\bf $I=2$ $\pi\pi$ potential in the HAL QCD method\\ with all-to-all propagators}
\author[1,2]{Yutaro Akahoshi}
\author[1,2]{Sinya Aoki}
\author[3,2]{Tatsumi Aoyama}
\author[2,4]{Takumi Doi}
\author[1,2]{Takaya Miyamoto}
\author[1,2]{Kenji Sasaki}
\affil[1]{\small Center for Gravitational Physics, Yukawa Institute for Theoretical Physics\\Kyoto University, Kyoto 606-8502, Japan}
\affil[2]{\small RIKEN Nishina Center (RNC), Saitama 351-0198, Japan}
\affil[3]{\small Institute of Particle and Nuclear Studies, High Energy Accelerator Research Organization(KEK)\\Tsukuba, Ibaraki 305-0801, Japan}
\affil[4]{\small RIKEN Interdisciplinary Theoretical and Mathematical Sciences Program (iTHEMS), Saitama 351-0198, Japan}

\begin{document}
\maketitle
\hspace{-0.6cm}\hrulefill
\begin{abstract}
In this paper, we perform the first application of the hybrid method (exact low modes plus stochastically estimated high modes) for all-to-all propagators
to the HAL QCD method.
We calculate the HAL QCD potentials in the $I=2$ $\pi\pi$ scattering
in order to see how statistical fluctuations of the potential behave under the hybrid method.
All of the calculations  are performed with the 2+1 flavor gauge configurations on a $16^3 \times 32$ lattice at the lattice spacing $a \approx 0.12$ fm and $m_{\pi} \approx 870$ MeV. It is revealed that statistical errors for the potential are enhanced by  stochastic noises introduced by the hybrid method,
which, however, are shown to be reduced by increasing the level of dilutions,
in particular, that of space dilutions.
From systematic studies, we obtain a guiding principle for a choice of dilution types/levels and
a number of eigenvectors to reduce noise contamination to the potential while keeping numerical costs reasonable.
We also confirm that we can obtain the scattering phase shifts for the $I=2$ $\pi\pi$ system
by the hybrid method within a reasonable numerical cost; these phase shifts are consistent with the result obtained with the conventional method.
The knowledge that we obtain in this study will become useful for the investgation of hadron resonances that
require quark annihilation diagrams such as the $\rho$ meson by the HAL QCD potential with the hybrid method.
\end{abstract}
\hrulefill

\section{Introduction}
Understanding hadronic resonances in terms of quantum chromodynamics (QCD) is one of the most challenging subjects in both particle and nuclear physics.
In order to study hadronic resonances in QCD, we have to analyze hadron--hadron scatterings
using non-perturbative methods such as lattice QCD,
where two methods have been employed so far.
One is L\"uscher's finite volume method (and the extensions thereof) \cite{luscher1,luscher2,luscher3},
which has been mainly applied to meson--meson systems (reviewed in Ref.~\cite{Briceno:2017max})  including resonant states such as $\rho$~\cite{Aoki:2007rd, luscher_rho1, luscher_rho2} and $\sigma$~\cite{luscher_sigma1, luscher_sigma2}.
The other one is the HAL QCD method\cite{hal1,hal2,hal3,hal4}, in which non-local but energy-independent potentials are constructed from the Nambu--Bethe--Salpeter (NBS) wave functions in lattice QCD, and
then physical observables such as scattering phase shifts are extracted
by solving the Schr\"odinger equations with the potentials.
The HAL QCD method has been applied to a wide range of hadronic systems\cite{hal_inouesan1,hal_inouesan2,hal_nemurasan,hal_muranosan1,hal_muranosan2,hal_sasakisan,hal_ikedasan,hal_etminansan,hal_yamadasan,hal_doisan,hal_miyamotosan}
including, in particular,
the candidate for the exotic state, $Z_c(3900)$~\cite{hal_zc1, hal_zc2}.
The consistency between L\"uscher's method and the HAL QCD method for two-baryon systems
is extensively studied 
in Refs.~\cite{Iritani:2016jie,Iritani:2017rlk,Iritani:2018zbt,Iritani:2018vfn}.
Recently, the first realistic calculation of baryon interactions
with nearly physical quark masses is performed in the HAL QCD method (a recent summary is given in Ref.~\cite{Doi_2018})
and $\Omega\Omega$ and $N\Omega$ systems are found to form
di-baryons located near the unitarity~\cite{hal_oo, Iritani:2018sra}.

Although the HAL QCD method is a powerful method to study hadron--hadron scatterings, we still do not have a mature way to treat scattering processes containing quark annihilation diagrams, due to the need for all elements of propagators (so-called all-to-all propagators).
Such quark annihilation diagrams typically appear in resonant channels; thus, establishing an efficient technique to treat them is an urgent issue for deeper understanding of hadronic resonances in the HAL QCD method.
A previous study\cite{hal_kawaisan}, which utilized the LapH method\cite{laph}
for the all-to-all propagator in the HAL QCD method,
revealed that non-locality in the definition of the Nambu--Bethe--Salpeter (NBS) wave function
introduced by the LapH smearing
enhances higher-order contributions in the derivative expansion of the potential,
so that the leading-order potential suffers large systematic uncertainties even at low energies.

In this study, we employ another technique to obtain the all-to-all propagator, namely the hybrid method\cite{hybrid}, with the HAL QCD method for the first time. This technique combines a low-mode spectral decomposition of the propagator together with stochastic estimations for remaining high modes.
In contrast with the LapH method, this method keeps the locality of quark fields and thus the locality of hadron operators in the HAL QCD potential.
In order to investigate how statistical errors of potentials and physical observables behave under the hybrid method,
we calculate potentials for the $I=2$ $\pi\pi$ S-wave scattering on gauge configurations at $m_{\pi} \approx 870$ MeV with all-to-all propagators.
Since the calculation of quark annihilation diagrams is not necessary,
this system is suitable to benchmark the hybrid method.
As a result, we find that the potential by the hybrid method  gives us reliable results as long as
statistical errors caused by stochastic noises are kept sufficiently small by
noise dilution techniques.
Our study also provides an optimal choice of parameters in the hybrid method to calculate the potential, which will be useful for future studies.

\vspace{5mm}
This paper is organized as follows. In Sect. 2, after briefly explaining the HAL QCD method,
we introduce the hybrid method and
discuss the application to the HAL QCD method.
Simulation details in this study are shown in Sect. 3.
As main results of our study,
we present systematic studies on behaviors of the HAL QCD potentials with the hybrid method in Sect. 4, and a comparison with previous results without all-to-all propagators in Sect. 5.
Our conclusion is given in Sect. 6.

\section{Method}
\subsection{HAL QCD method}
A basic quantity in the HAL QCD method is the Nambu--Bethe--Salpeter (NBS) wave function,
which is given for the $I=2$ two-pion system as
\begin{equation}
  \psi_{W}({\bf r}) =  \sum_{\bf x} \langle 0| \pi^{+}({\bf r+x},0) \pi^{+}({\bf x},0) |\pi^{+}\pi^{+};{\bf k} \rangle,
\end{equation}
where $|\pi^{+}\pi^{+};{\bf k} \rangle$ is an asymptotic state of the elastic $\pi^{+}\pi^{+}$ S-wave system in the center-of-mass frame with a relative momentum ${\bf k}$ and an energy $W = 2 \sqrt{m_{\pi}^2 + k^2}$, $\quad k=\vert{\bf k}\vert$, and
the positively charged pion operator is given as $\pi^{+}({\bf x},t) = \bar d({\bf x},t) \gamma_5 u({\bf x},t)$
with  up and down quark fields $u({\bf x},t)$ and $d({\bf x},t)$.

For $r =\vert {\bf r}\vert > R$ with an interaction range $R$,
the NBS wave function satisfies  the Helmholtz equation as
\begin{eqnarray}
  (\nabla^2+k^2)\psi_{W}({\bf r}) &=& 0,
\end{eqnarray}
and its radial part with angular momentum $l$ behaves as
 \begin{eqnarray}
  \psi_W^l(kr) &=& C e^{i\delta_l(k)} \frac{\sin (kr-\frac{l \pi}{2} + \delta_l(k))}{kr},
\label{eq:Asymptotic}
\end{eqnarray}
where $\delta_l(k)$ is the scattering phase shift corresponding to the phase of  the S-matrix
constrained by the unitarity\cite{hal2,hal_nbs_ls} and $C$ is a constant.
 An energy-independent and non-local potential $U({\bf r},{\bf r'})$ is defined from
\begin{equation} \label{eq:method1}
  \frac{1}{2 \mu}(\nabla^2 + k^2) \psi_{W}({\bf r}) = \int d^3 {\bf r'}\, U({\bf r},{\bf r'})\psi_{W}({\bf r'}),
\end{equation}
where $\mu= m_{\pi}/2$ is the reduced mass of the two-pion.
Thanks to Eq.~(\ref{eq:Asymptotic}), this potential
is faithful to the scattering phase shifts by construction,
but the potential depends on a scheme such as
the choice of operators in the definition of the NBS wave function\cite{hal2,hal4}.
In practice, we deal with the non-locality of the potential  in the derivative expansion as
\begin{equation}
  U({\bf r},{\bf r'}) = (V^{\rm LO}(r) + V^{\rm N^2LO}(r) \nabla^2 + {\mathcal O}(\nabla^4)) \delta({\bf r-r'}),
\end{equation}
whose convergence property depends on the scheme of the potential.
In the previous study\cite{hal_kawaisan},
it was found that
non-locality of the operator in the NBS wave function
was a major factor that governs the non-locality of the potential:
A scheme with non-local (smeared) operators
leads to a potential with large non-locality and inclusion of $V^{\rm N^2LO}(r)$ is required to reproduce the $I=2$ $\pi\pi$ scattering phase shifts even at low energies,
while the scheme with  local operators leads to the potential with small non-locality and
$V^{\rm LO}(r)$ alone is found to be sufficient.
In this paper, we employ the scheme with local operators and calculate the potential at the leading order (LO), $V^{\rm LO}(r)$.

We define a four-point correlation function as
\begin{equation}
  F({\bf r},t-t_0) = \sum_{\bf x} \langle \pi^{+}({\bf r+x},t) \pi^{+}({\bf x},t) {\mathcal J}_{\pi^{+}\pi^{+}}(t_0) \rangle,
\end{equation}
where a source operator ${\mathcal J}_{\pi^{+}\pi^{+}}(t_0)$ creates $I=2$  $\pi^{+}\pi^{+}$ scattering states in the $A_1^{+}$ representation.
In this paper, we employ a bi-local operator in which each pion operator is projected to zero momentum:
\begin{equation} \label{eq:method9}
  {\mathcal J}_{\pi^{+}\pi^{+}}(t_0) = \sum_{\bf y,z} \pi^{-}({\bf y},t_0)\pi^{-}({\bf z},t_0).
\end{equation}
We then normalize it as
\begin{equation}
  R({\bf r},t) \equiv \frac{F({\bf r},t)}{(e^{-m_{\pi} t})^2},
\end{equation}
which can be written as
\begin{eqnarray} \label{eq:method3}
  R({\bf r},t) = \sum_n B_n \psi_{W_n}({\bf r}) e^{- \Delta W_n t} + ({\rm inelastic\ contributions}),
\end{eqnarray}
where $\Delta W_n = W_n - 2 m_{\pi}$ and $W_n = 2\sqrt{k_n^2 + m_{\pi}^2}$ is the energy of the $n$th elastic state.
Since
\begin{equation} \label{eq:method4}
  (\Delta W_n)^2 = 4 k_n^2 - 4 m_{\pi} \Delta W_n,
\end{equation}
the time-dependent HAL QCD method reads\cite{hal3}
\begin{equation}
  \left[ \frac{\nabla^2}{m_{\pi}} -\frac{\partial}{\partial t} + \frac{1}{4m_{\pi}} \frac{\partial^2}{\partial t^2} \right] R({\bf r},t) = \int d^3{\bf r'} U({\bf r},{\bf r'}) R({\bf r'},t),
  \label{eq:t-dep_hal}
\end{equation}
as long as $t$ is large enough to suppress inelastic contributions in $R({\bf r},t)$.
Thus the LO potential is given by
\begin{equation}
  V^{\rm LO}(r) = \frac{\left[ \frac{\nabla^2}{m_{\pi}} -\frac{\partial}{\partial t} + \frac{1}{4m_{\pi}} \frac{\partial^2}{\partial t^2} \right] R({\bf r},t)}{R({\bf r},t)}.
    \label{eq:t-dep_hal:LO}
\end{equation}

\subsection{The hybrid method for all-to-all propagators}
In this paper, we employ the hybrid method for all-to-all propagators\cite{hybrid},
where the dominant part of the quark propagator is represented by low eigenmodes of the Dirac operators while the remaining contributions from high eigenmodes are estimated stochastically.

The quark propagator can be expressed in terms of eigenmodes of the Hermitian Dirac operator $H=\gamma_5 D$ as
\begin{equation} \label{eq:method5}
  D^{-1}(x,y) = \sum_{i=0}^{N-1} \frac{1}{\lambda_i} v^{(i)}(x) \otimes v^{(i)}(y)^{\dag} \gamma_5,
\end{equation}
where $\lambda_i$ and $v^{(i)}$ are the $i$th eigenvalue and eigenvector of $H$, respectively, and $N$ is the total number of eigenmodes. Color and spinor indices are implicit here.
We approximate the propagator using the low-lying $N_{\rm eig}$ eigenmodes as
\begin{equation}
  D_0^{-1}(x,y) = \sum_{i=0}^{N_{\rm eig}-1} \frac{1}{\lambda_i} v^{(i)}(x) \otimes v^{(i)}(y)^{\dag} \gamma_5,
\end{equation}
which is expected to be a reasonable approximation for pions at low energies.

The rest,
\begin{equation}
M^{-1} \equiv   D^{-1} - D_0^{-1} = H^{-1} {\mathcal P}_1 \gamma_5,
\quad  {\mathcal P}_1 \equiv
{\bf 1} - \sum_{i=0}^{N_{\rm eig}-1}
v^{(i)} \otimes v^{\dag (i)},
\end{equation}
can be estimated stochastically as
\begin{equation}
  M^{-1}(x,y) = \langle \hspace{-0.5mm} \langle \psi(x) \otimes \eta(y)^{\dag} \rangle \hspace{-0.5mm} \rangle \gamma_5,
\end{equation}
where a noise vector $\eta$ satisfies
\begin{eqnarray}
  \langle \hspace{-0.5mm} \langle \eta_{a \alpha}(x) \otimes \eta_{b \beta}(y)^{\dag} \rangle \hspace{-0.5mm} \rangle &=& \delta_{a,b} \delta_{\alpha,\beta} \delta_{x,y} \label{eq:method6} \\
  |\eta_{a \alpha}(x)|^2 &=& 1\ ({\rm no \ summation})
\end{eqnarray}
for color indices $a,b$ and spinor indices $\alpha,\beta$,
and $\psi$ is the solution of $ H \cdot \psi = {\mathcal P}_1 \eta$.
The symbol $\langle \hspace{-0.5mm} \langle \  \rangle \hspace{-0.5mm} \rangle$ indicates an expectation value over probability distribution of noise vectors.

\begin{figure}[bp]
  \centering
  \begin{tabular}{lr}
    \hspace{-7mm}
    \begin{minipage}{130pt}
      \includegraphics[width=130pt,clip]{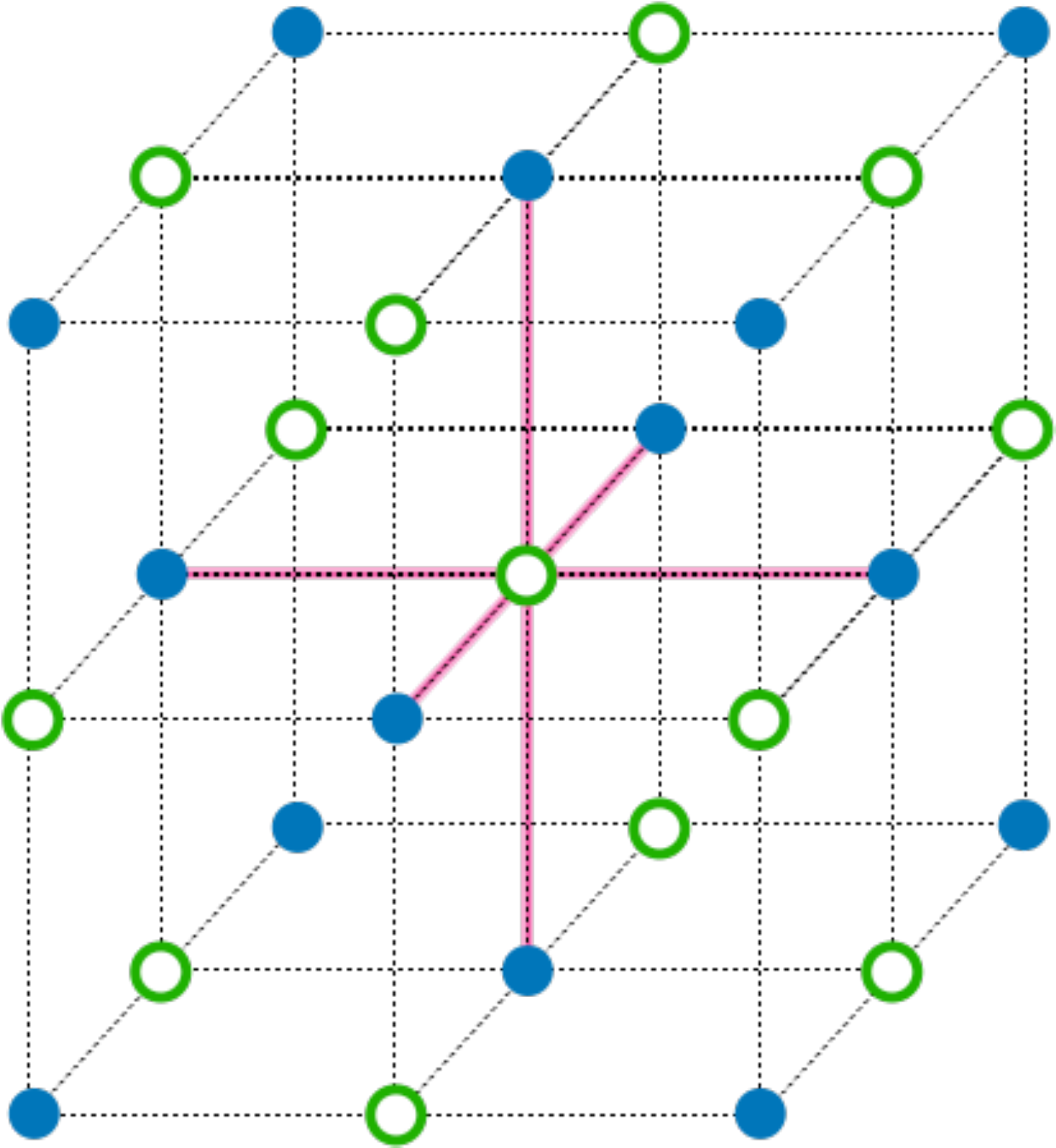}
    \end{minipage} & \hspace{15mm}
    \begin{minipage}{130pt}
      \includegraphics[width=130pt,clip]{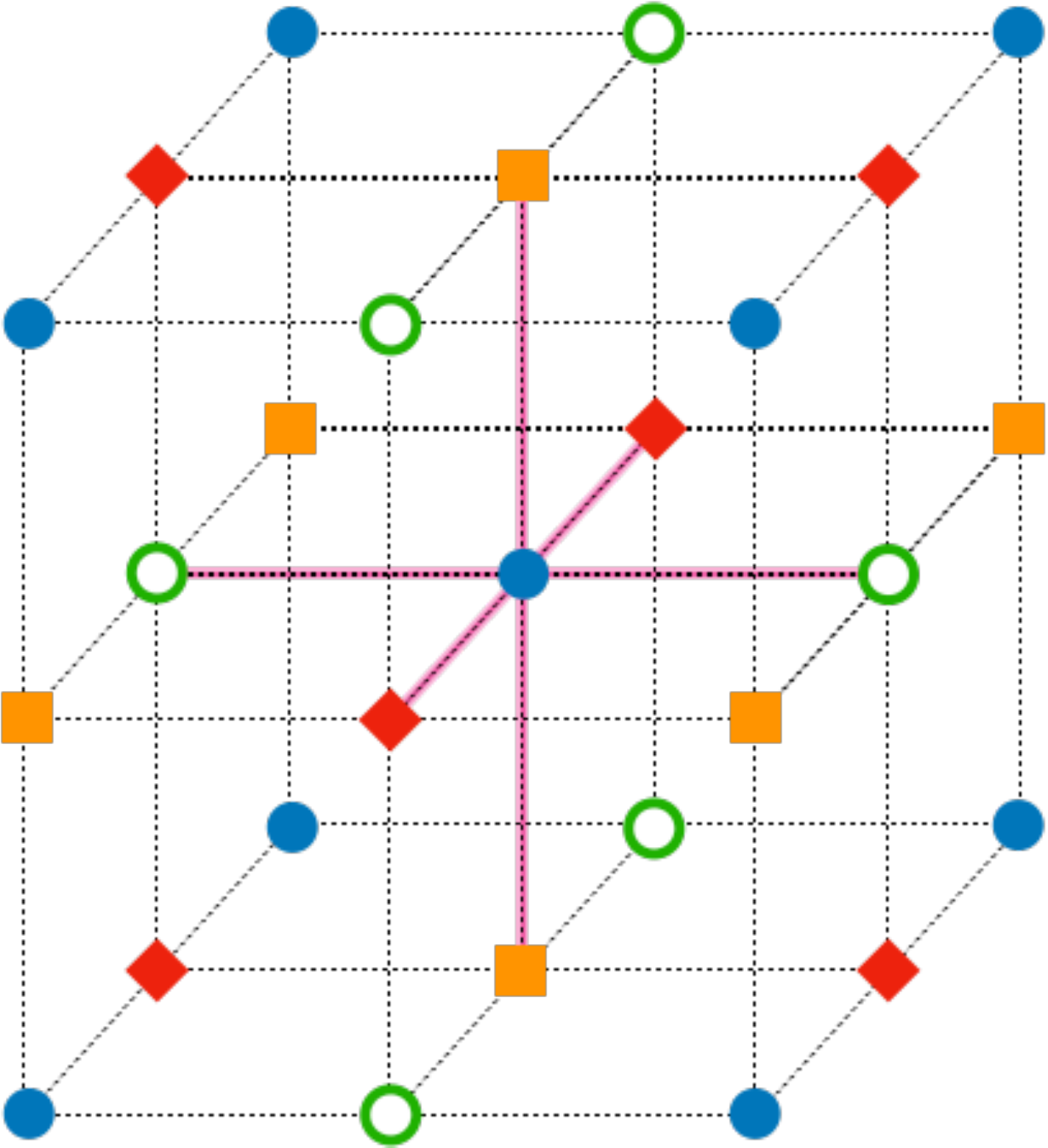}
    \end{minipage}
  \end{tabular}
  \caption{Schematic figures of $s2$ dilution (left) and
  $s4$ dilution (right). Sites with different symbols belong to different diluted noises.
  Pink lines connect sites that are referred to in the calculation of the Laplacian at the central point.}
  \label{fig:spacedils}
\end{figure}

In practice, the exact noise average is approximated by the finite sum, together with the variance  reduction by diluted noises as
\begin{equation} \label{eq:method8}
  M^{-1}(x,y) \approx \frac{1}{N_{\rm r}} \sum_{r=0}^{N_{\rm r}-1} \sum_{i=0}^{N_{\rm dil}-1} \psi_{[r]}^{(i)}(x) \otimes \eta_{[r]}^{(i)}(y)^{\dag} \gamma_5,
 \end{equation}
where $N_r$ is the total number of noises,
$N_{\rm dil}$ is the total number of dilution, and
$\psi_{[r]}^{(i)}$ is the solution of $ H\cdot  \psi_{[r]}^{(i)}= {\mathcal P}_1 \eta_{[r]}^{(i)}$.
Using diluted noises,
parts of statistical errors that appear in off-diagonal elements of the approximation of Eq.~(\ref{eq:method6}) become explicitly zero, and therefore the variance reduction is achieved.
In our study, color and spinor are fully diluted, while several types of dilutions are employed for time and space. The time coordinate is either fully or $J$-interlace diluted by the noises as
\begin{equation}
\eta^{(i)}({\bf x}, t) \not= 0, \quad  \mbox{if $t= i$ mod $J$},
\end{equation}
where $i$ runs from 0 to  $J-1$; thus $J=N_t$ corresponds to the full dilution.
On the other hand, no dilution, $s2$ (even/odd) dilution, or $s4$ dilution is employed for spatial coordinates.
Noise vectors for the $s2$ dilution are given by
\begin{equation}
  \eta^{(i)}(x,y,z, t) \neq 0, \quad \mbox{if  $x+y+z  =i$ mod 2},
\end{equation}
while those for the $s4$ dilution read
\begin{eqnarray}
\left\{
\begin{array}{ccc}
 \eta^{(0)} \neq 0 & \mbox{if $(n_x,n_y,n_z)$ = (even,even,even) or (odd,odd,odd)}    \\
  \eta^{(1)} \neq 0 & \mbox{if $(n_x,n_y,n_z)$ = (odd,even,even) or (even,odd,odd)}     \\
  \eta^{(2)} \neq 0 & \mbox{if $(n_x,n_y,n_z)$ = (even,odd,even) or (odd,even,odd)}     \\
  \eta^{(3)} \neq 0 & \mbox{if $(n_x,n_y,n_z)$ = (odd,odd,even) or (even,even,odd)}     \\
\end{array}
\right. .
\end{eqnarray}
These space dilutions are schematically shown in Fig.~\ref{fig:spacedils}.
These dilutions are chosen so as to minimize the off-diagonal noises from the neighbor sites.
In particular, since the Laplacian is evaluated by the second-order difference
using one central site and six next-nearest-neighbor sites,
we maximize the number of independent noises assigned to these seven sites.
Thus, the total number of dilutions becomes $N_{\rm dil} = 3\times 4\times J\times 2^{s/2}$ with $s=0$ (no dilution), $s=2$ ($s2$), or $s=4$ ($s4$).

In total, the all-to-all propagator in the hybrid method is given by
\begin{equation}
  D^{-1} \approx \frac{1}{N_{\rm r}} \sum_{r=0}^{N_{\rm r}-1} \sum_{i=0}^{N_{\rm hl}-1} u_{[r]}^{(i)} \otimes w_{[r]}^{\dag (i)} \gamma_5,
\end{equation}
where the hybrid lists $u_{[r]}^{(i)},w_{[r]}^{(i)}$ are defined as
\begin{eqnarray}
  w_{[r]}^{(i)} &=& \{ \frac{v^{(0)}}{\lambda_0}, \cdots , \frac{v^{(N_{\rm eig}-1)}}{\lambda_{N_{\rm eig}-1}},\eta_{[r]}^{(0)}, \cdots ,\eta_{[r]}^{(N_{\rm dil}-1)} \} \\
  u_{[r]}^{(i)} &=& \{ v^{(0)}, \cdots , v^{(N_{\rm eig}-1)}, \psi_{[r]}^{(0)}, \cdots ,\psi_{[r]}^{(N_{\rm dil}-1)} \}
\end{eqnarray}
with $N_{\rm hl} = N_{\rm eig} + N_{\rm dil}$.
The HAL QCD potential with the local operator can be constructed by using this all-to-all propagator.

\subsection{Correlation functions in the hybrid method}
The two-point correlation function for the charged pion,
\begin{equation}
  C(t) = \sum_{{\bf x,y},t_0} \langle \pi^{+}({\bf x},t+t_0) \pi^{-}({\bf y},t_0) \rangle,
\end{equation}
is expressed as
\begin{eqnarray}
C(t)
    &=& - \frac{1}{N_{\rm r}^2}\sum_{t_0} \sum_{r,s} \sum_{i,j} O^{(j,i)}_{[s,r]}(t+t_0) O^{(i,j)}_{[r,s]} (t_0),
\end{eqnarray}
where
\begin{equation} \label{eq:appex2}
  O^{(i,j)}_{[r,s]}(t) = \sum_{{\bf x}} O^{(i,j)}_{[r,s]}({\bf x},t), \quad
 O^{(i,j)}_{[r,s]}({\bf x},t) \equiv  w^{\dag (i)}_{[r]} ({\bf x},t) \cdot u^{(j)}_{[s]}({\bf x},t),
\end{equation}
and the dot $\cdot$ indicates an inner product in color and spinor indices.

The four-point correlation function defined by
\begin{equation}
  F({\bf r},t) = \sum_{\bf x,y_1,y_2,t_0} \langle \pi^{+}({\bf r+x},t+t_0) \pi^{+}({\bf x},t+t_0) \pi^{-}({\bf y_1},t_0)\pi^{-}({\bf y_2},t_0) \rangle
\end{equation}
leads to
\begin{eqnarray}
 F({\bf r},t) &=&
 \frac{1}{N_{\rm r}^4} \sum_{{\bf x},t_0,r,s,p,q} \sum_{i,j,k,l} \left[O_{[r,s]}^{(i,j)}({\bf r+x},t+t_0) O_{[s,r]}^{(j,i)}(t_0) O_{[p,q]}^{(k,l)}({\bf x},t+t_0) O_{[q,p]}^{(l,k)}(t_0) \right. \nonumber \\
 &-& \left. O_{[r,s]}^{(i,j)}({\bf r+x},t+t_0) O_{[s,p]}^{(j,k)}(t_0) O_{[p,q]}^{(k,l)}({\bf x},t+t_0) O_{[q,r]}^{(l,i)}(t_0) \right] .
\end{eqnarray}

\section{Simulation details}
We employ the same 2+1 flavor QCD ensemble as the previous study\cite{hal_kawaisan},
generated by the JLQCD and CP-PACS Collaborations\cite{jlqcd,cppacsjlqcd}
on a $16^3 \times 32$ lattice  with an Iwasaki gauge action\cite{iwasakiaction} at $\beta=1.83$ and a non-perturbatively improved Wilson-clover action\cite{cloveraction} at $c_{SW} = 1.7610$ and hopping parameters $(\kappa_{ud},\kappa_s) = (0.1376,0.1371)$,
which correspond to
the lattice spacing $a = 0.1214$ fm and the pion mass $m_{\pi} \approx 870$ MeV.

In addition to the local quark source, we use the smeared quark source~\cite{Iritani:2016jie},
$q_s({\bf x},t) =\sum_{\bf y} f({\bf x} -{\bf y}) q({\bf y},t)$ with the Coulomb gauge fixing,
where
\begin{equation}
  f \left( {\bf x}  \right) = \begin{cases}
    a e^{-b |{\bf x }|} & ( \ 0 < |{\bf x }| < (L-1)/2 \ ) \\
    1 & ( \ |{\bf x }| = 0 \ ) \\
    0 & ( \ |{\bf x }| \geq (L-1)/2 \ )
  \end{cases}
\end{equation}
with $a=1.0,b=0.47$ in lattice units. Note that,  regardless of the type of quark sources, all calculations are made with wall sources at the hadron level (see Eq.(\ref{eq:method9})).
The periodic boundary condition is used in all directions.

The setups for the hybrid method in this study are presented in Table~\ref{tab:result2}.
In this study, we use a single noise vector for each propagator ($N_{\rm r} = 1$), and the noise vectors are generated by $Z_4$ random noises.
Statistical errors are estimated by the jackknife method with a bin-size of 1 except for the case5a, where the bin-size is 6.
\begin{table}[tbp]
  \caption{Setups for the hybrid method. $N_{\rm eig}$ is the number of low eigenmodes for the all-to-all propagator,
      and the number of the noise vector for high eigenmodes is 1 for all cases. Color and spinor dilutions are always used.}
  \label{tab:result2}
  \centering
  \begin{tabular}{c|ccccc}
     & time dilution & space dilution & $N_{\rm eig}$ & Source & $N_{\rm conf}$ \\ \hline \hline
    case1 & full & none & 100 & point & 20 \\
    case2 & full & $s2$ (even/odd) & 100 & point & 20 \\
    case3 & 16-interlace &  $s2$ (even/odd) & 100 & point & 20 \\
    case4 & 16-interlace &  $s2$ (even/odd) & 100 & smear & 20 \\
    case5 & 16-interlace & $s4$ & 100 & smear & 20 \\
    case5a & 16-interlace & $s4$ & 100 & smear & 60 \\
    case6 & 16-interlace& $s2$ (even/odd) & 200 & point & 20 \\
    case7 & 16-interlace& $s2$ (even/odd) & 484 & smear & 20 \\
    \hline
  \end{tabular}
\end{table}

\begin{figure}[htbp]
  \begin{tabular}{lr}
    \hspace{-7mm}
    \begin{minipage}{230pt}
      \centering
      \includegraphics[width=230pt,clip]{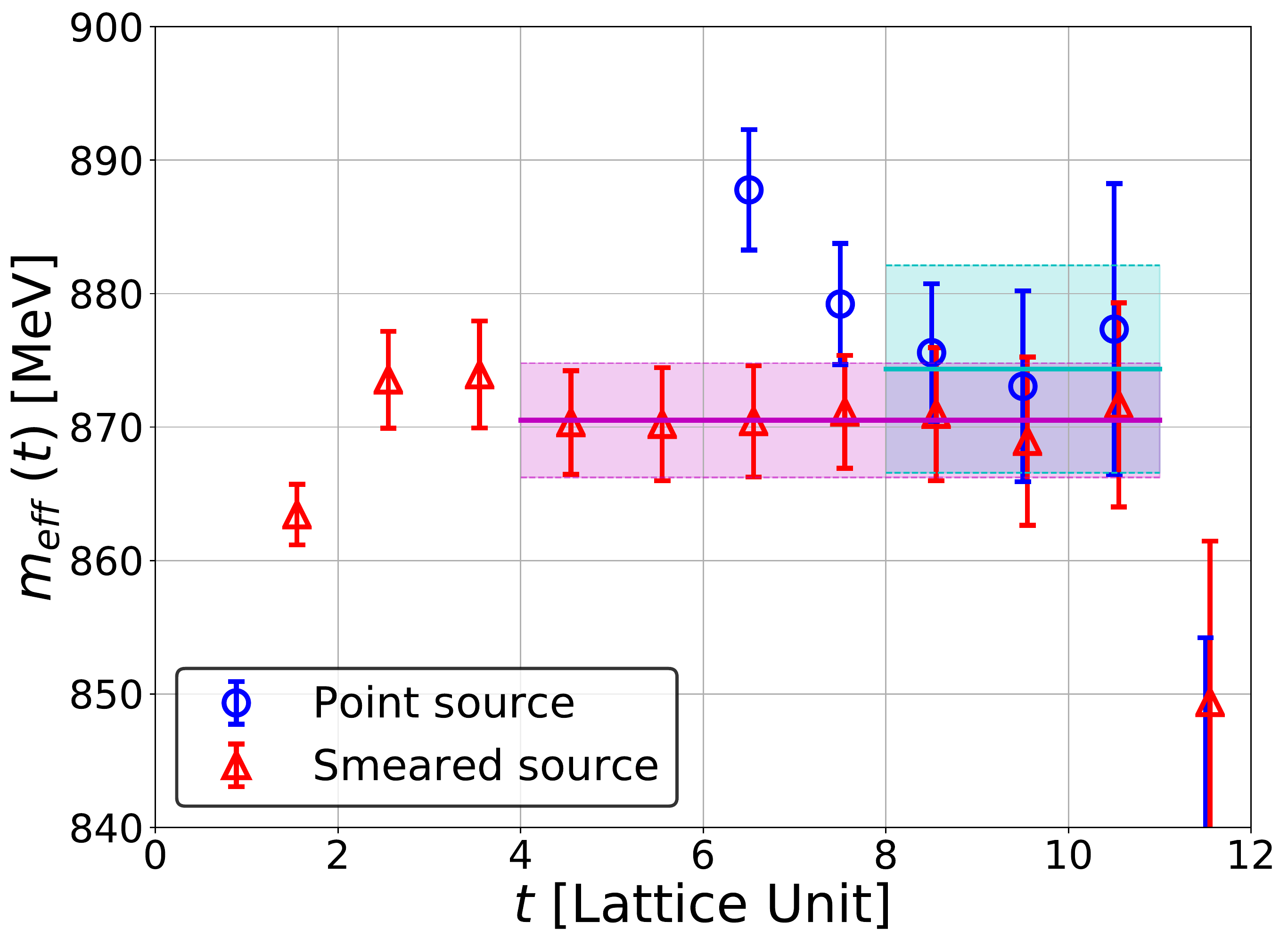}
    \end{minipage}
    & \hspace{-5mm}
    \begin{minipage}{230pt}
      \centering
      \includegraphics[width=230pt,clip]{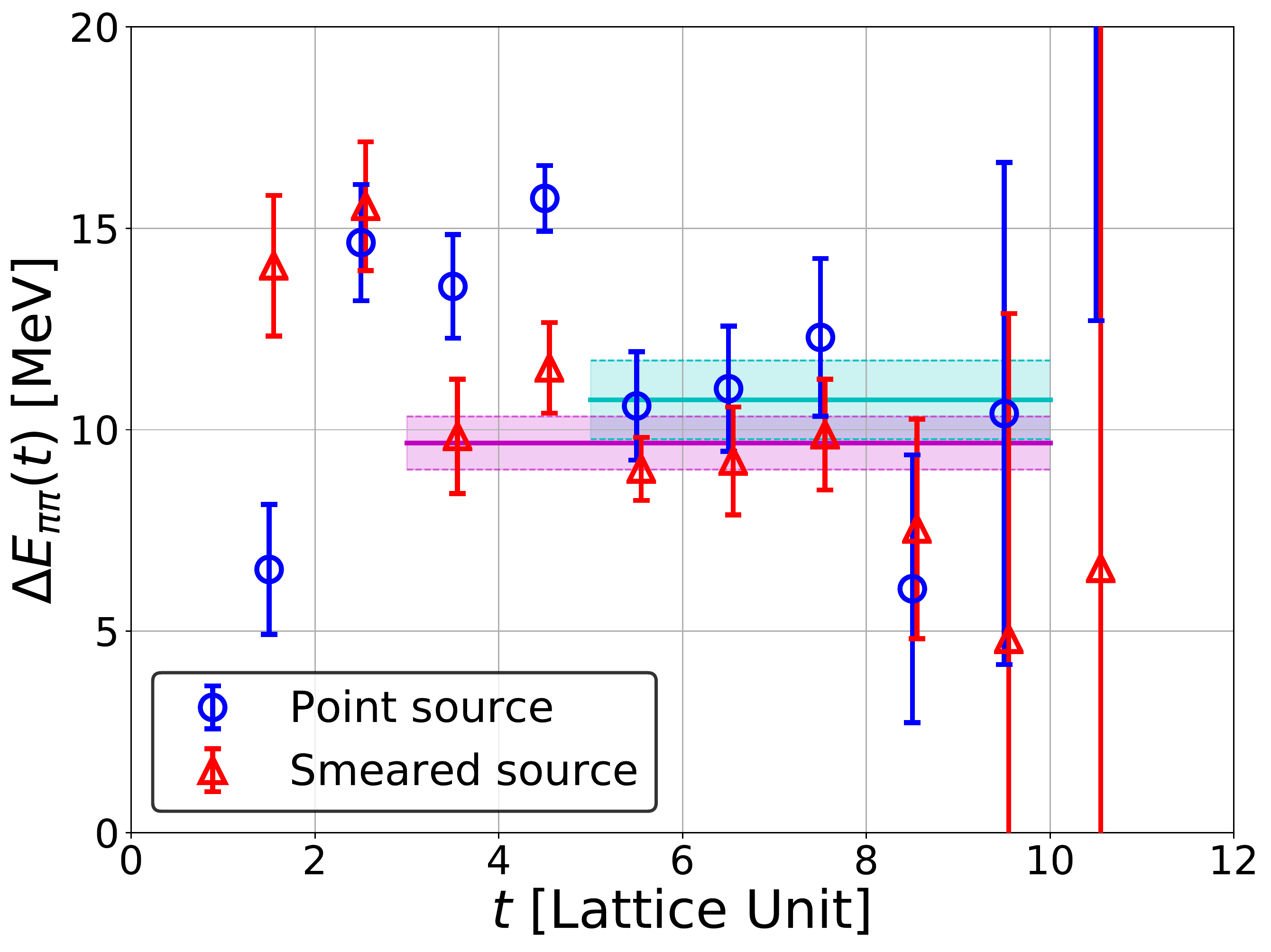}
    \end{minipage}
  \end{tabular}
  \caption{(Left) Effective mass of the single pion from the point source(blue circles) and the smeared source (red triangles).
  (Right) Effective energy shift of two pions,  $\Delta E_{\pi\pi} = E_{\pi\pi} - 2m_\pi$, from
  the point(blue circles)  and smeared(red triangles) sources. Cyan and magenta solid lines with bands represent central values and statistical errors from fits to point and smeared data in these intervals, respectively.}
  \label{fig:result11}
\end{figure}

Figure~\ref{fig:result11} (left) shows the effective masses of the single pion,
where
effective masses are calculated by solving the following equation,
\begin{equation}
  \frac{C(t)}{C(t+1)} = \frac{\cosh \left( m_{\rm eff}(t+1/2) (t - T/2) \right)}{\cosh \left( m_{\rm eff}(t+1/2) (t + 1 - T/2) \right)}.
\end{equation}
Note that we use a half-integer time convention for $m_{\rm eff}$,
whose convention is also used
for the effective energy shift.
  We find that the result from the smeared source reaches the plateau at earlier time slices than
  the one from the point source.
The fit to smeared data at $t=$4--11 gives $m_\pi = 870(4)$ MeV,
 while the fit to point data at $t=$8--11 leads to $m_{\pi} = 874(8)$ MeV,
 and both agree with 870 MeV in the previous study\cite{hal_kawaisan}.

Effective energy shifts  for two pions from smeared and point sources are plotted in Fig.~\ref{fig:result11} (right),
where the energy shift is defined by
$\Delta E_{\pi\pi} = E_{\pi\pi} - 2m_\pi$ with the two-pion energy $E_{\pi\pi}$.
In our setup, an energy gap between the ground and the first excited states is estimated to be $\Delta E_1 \sim 420$ MeV in the non-interacting case; thus ground state saturation is expected to be achieved at roughly $t \sim \frac{1}{\Delta E_1} \sim 3.8$ in lattice units.
Fits give
$\Delta E_{\pi\pi} = 9.7(0.7)$ MeV ($t=$ 3--10) from the smeared source, and
$\Delta E_{\pi\pi} = 11(1)$ MeV ($t=$ 5--10) from the point source,
which are consistent with $\Delta E_{\pi\pi} = 14(5)$ MeV in the previous study\cite{hal_kawaisan}.
These results suggest that $R({\bf r}, t)$ is dominated by the ground state actually at $t\ge 5$, so that
the leading-order potential obtained by the time-dependent HAL QCD method becomes reliable at low energies.

\section{The hybrid method and the HAL QCD potential}
In this section, we show how statistical errors of the HAL QCD potential for the $I=2$ $\pi\pi$ system
depend on various setups of the hybrid method.
We mainly discuss data at $t=6$, which is sufficiently large for the elastic state domination as discussed in the previous section.
In the following, we use a quartet (time dilution, space dilution, $N_{\rm eig}$, source type) to specify calculation setups.

\begin{figure}[htbp]
  \centering
  \includegraphics[width=390pt,clip]{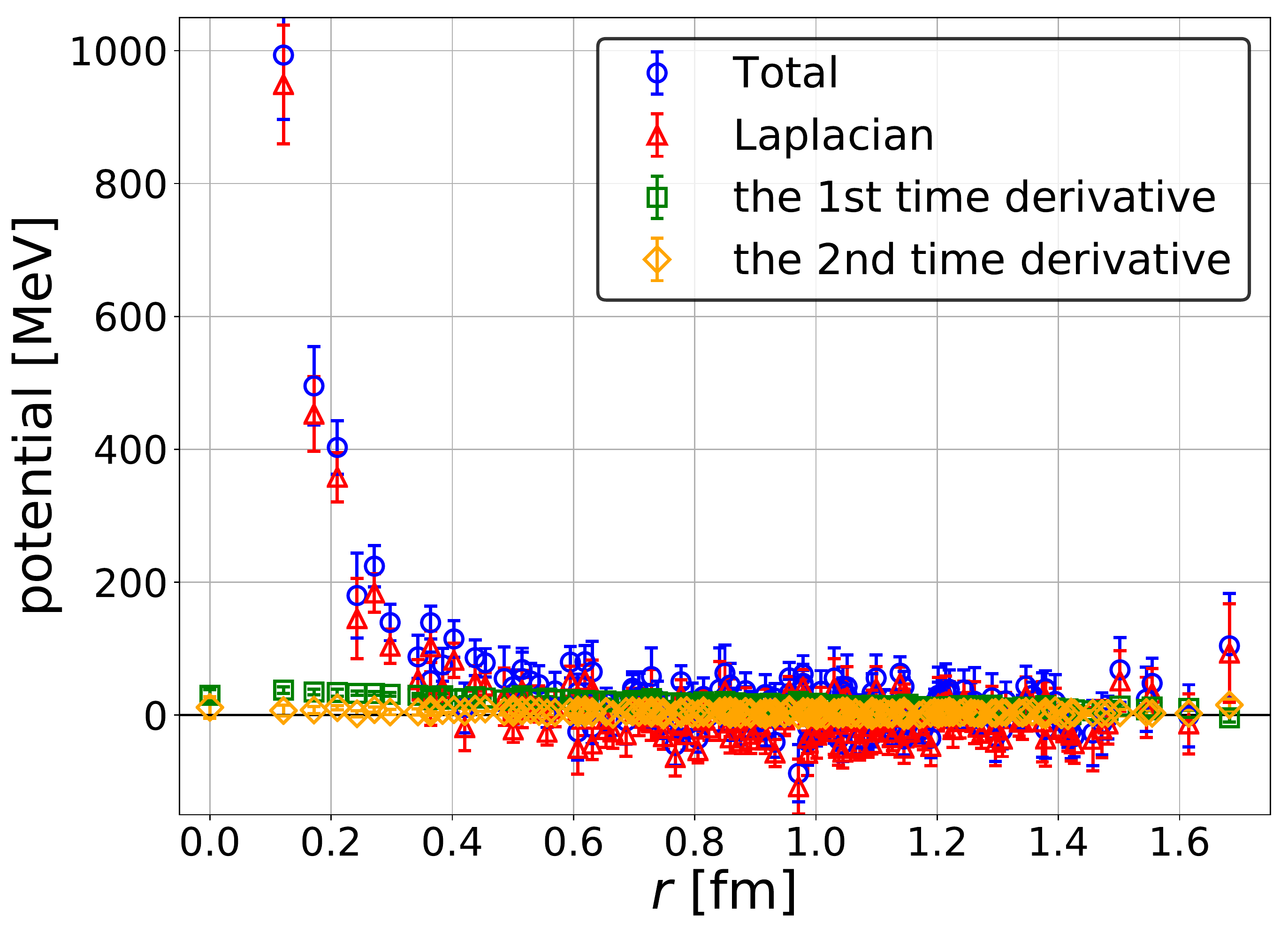}
  \caption{The potential from the hybrid method with
    case1 (full, none, 100, point) at $t=6$ (blue circles), together with its breakdown to three contributions, the Laplacian (red triangles),
    the first time derivative (green squares), and the second time derivative (yellow diamonds).
  }
  \label{fig:result2}
\end{figure}

Figure~\ref{fig:result2} shows the $I=2$ $\pi\pi$ potential at $t=6$ from the point source with the full time dilution (case1), together with its decomposition into the first (Laplacian), the second (the first time derivative) and the third (the second time derivative) terms on the right-hand side of Eq.~(\ref{eq:t-dep_hal:LO}).
Although the bulk behavior of the potential agrees with the previous result\cite{hal_kawaisan},
the potential obtained by the hybrid method suffers from much larger statistical errors,
which mainly come from the Laplacian term.

\subsection{Dilution in spatial directions}
\begin{figure}[htbp]
  \centering
  \begin{tabular}{lr}
  \hspace{-7mm}
  \begin{minipage}{230pt}
    \includegraphics[width=230pt,clip]{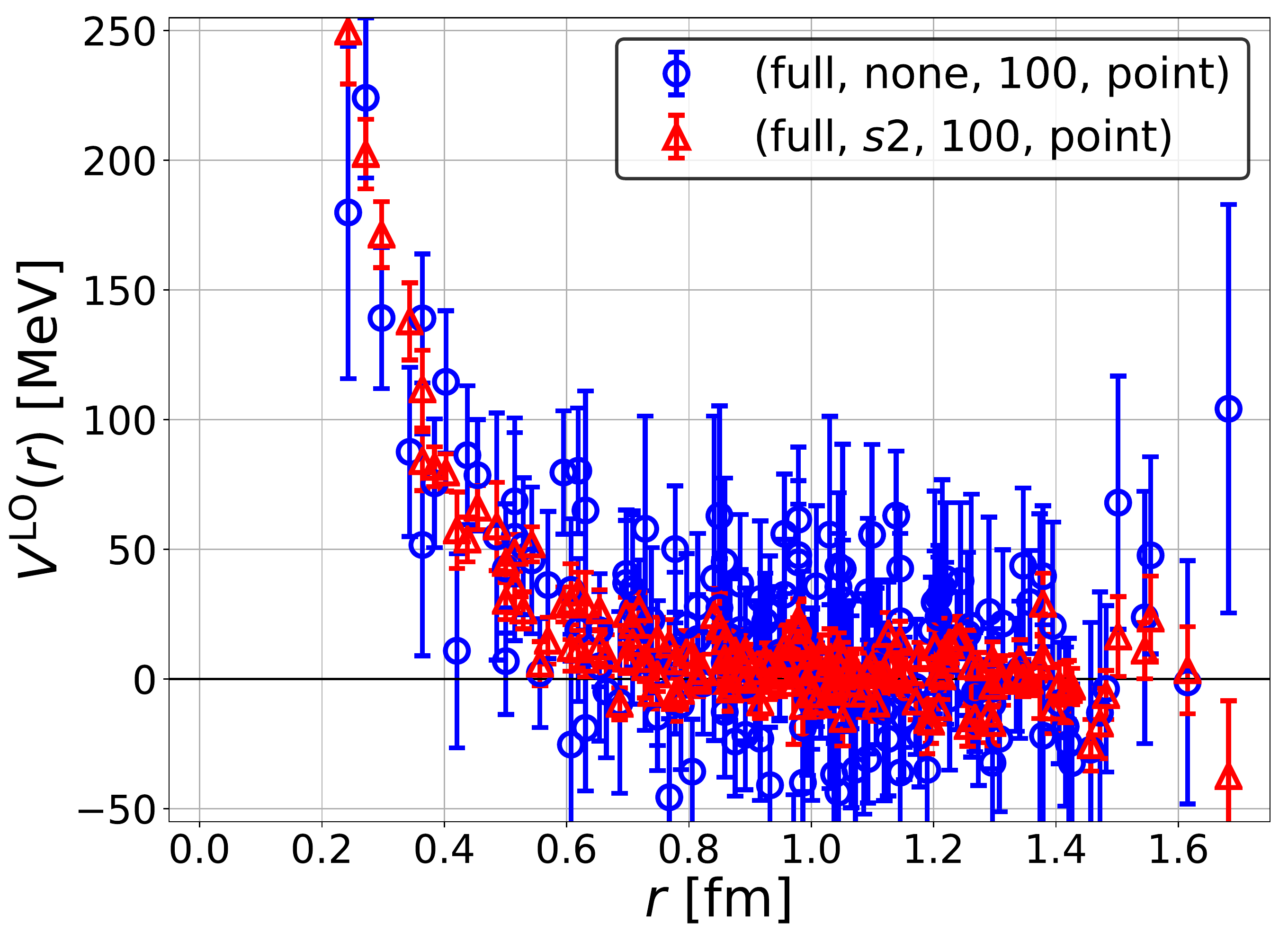}
  \end{minipage} & \hspace{-5mm}
  \begin{minipage}{230pt}
    \includegraphics[width=230pt,clip]{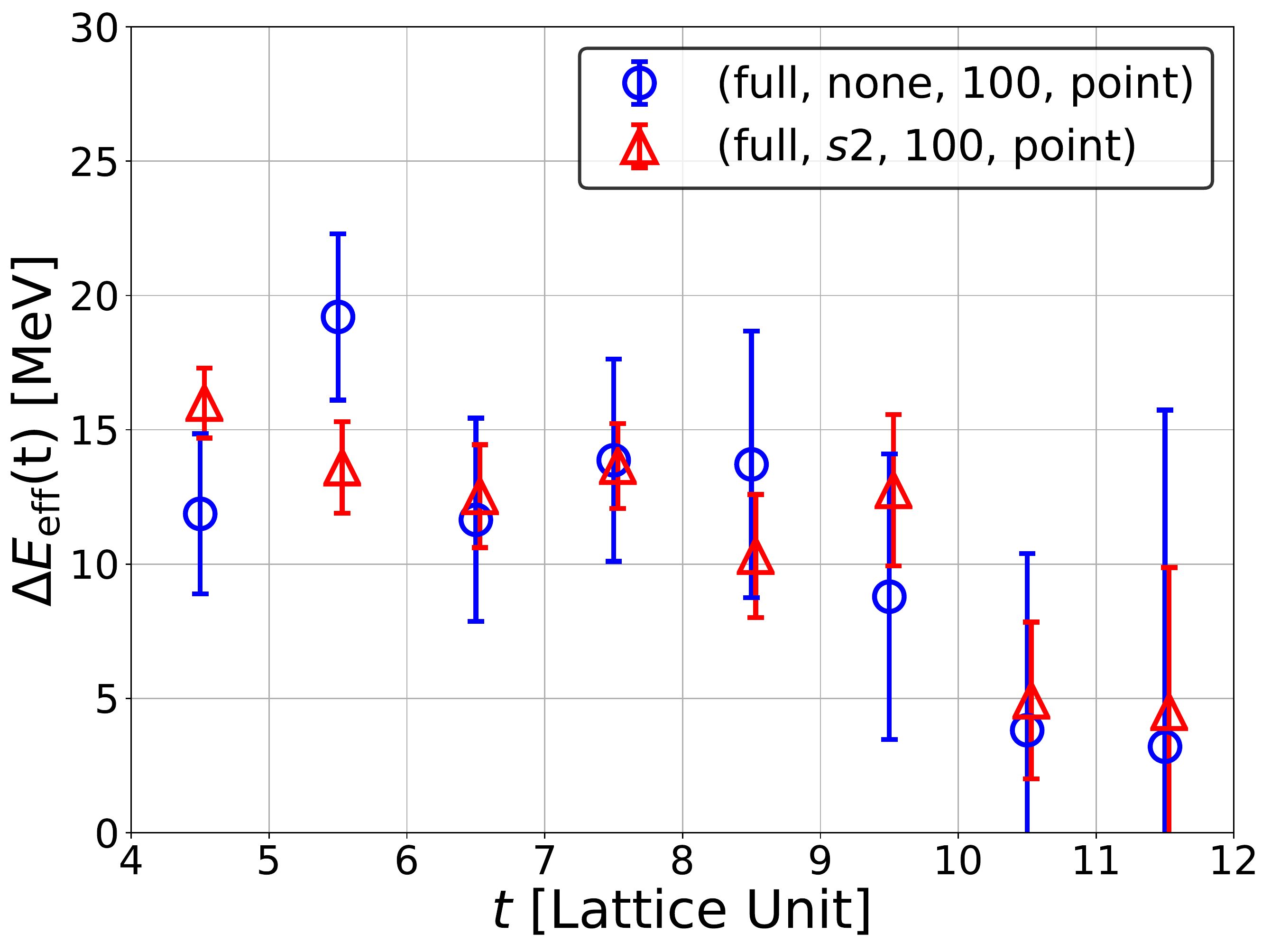}
  \end{minipage}
  \end{tabular}
  \caption{(Left) Dependence of the potential on space dilutions at $t=6$.
    (Right) Dependence of the effective energy shift on space dilutions.
    Data with blue circles and red triangles are obtained from case1 (no space dilution)
      and case2 ($s2$ space dilution), respectively.}
  \label{fig:result3}
\end{figure}

In order to reduce noise contamination in the Laplacian term, we introduce the $s2$ dilution
for spatial coordinates (case2) in addition to the full time dilution.
Shown in Fig.\ref{fig:result3} (left) is the corresponding result
together with that from the full time dilution only (case1) for comparison.
As can be seen from the figure, the statistical errors of the potential are much more reduced by the $s2$ dilution for spatial coordinates.
Although the numerical cost in case2 becomes approximately twice as large as that in case1, the statistical errors decrease by a factor of $\sim 3$,
and therefore we can conclude that the space dilution actually reduces the statistical noise.

On the other hand, as seen in Fig.\ref{fig:result3} (right), which compares the effective energy shifts $\Delta E_{\rm eff}$ between case1 and case2, their difference is moderate in contrast to the potentials, while the
magnitude of errors becomes smaller by space dilutions.
This observation suggests that the summation over spatial coordinates for the calculation of the energy shift reduces noise contamination, thanks to cancellation among different spatial points.
We thus conclude that
noise reduction by dilutions in spatial directions is more important for the HAL QCD potential than for the energy shift calculation, as the potential is extracted from spatial as well as temporal dependences of correlation functions.

\subsection{Dilution in a temporal direction}
\begin{figure}[htbp]
  \centering
  \begin{tabular}{lr}
    \hspace{-7mm}
    \begin{minipage}{230pt}
      \includegraphics[width=230pt,clip]{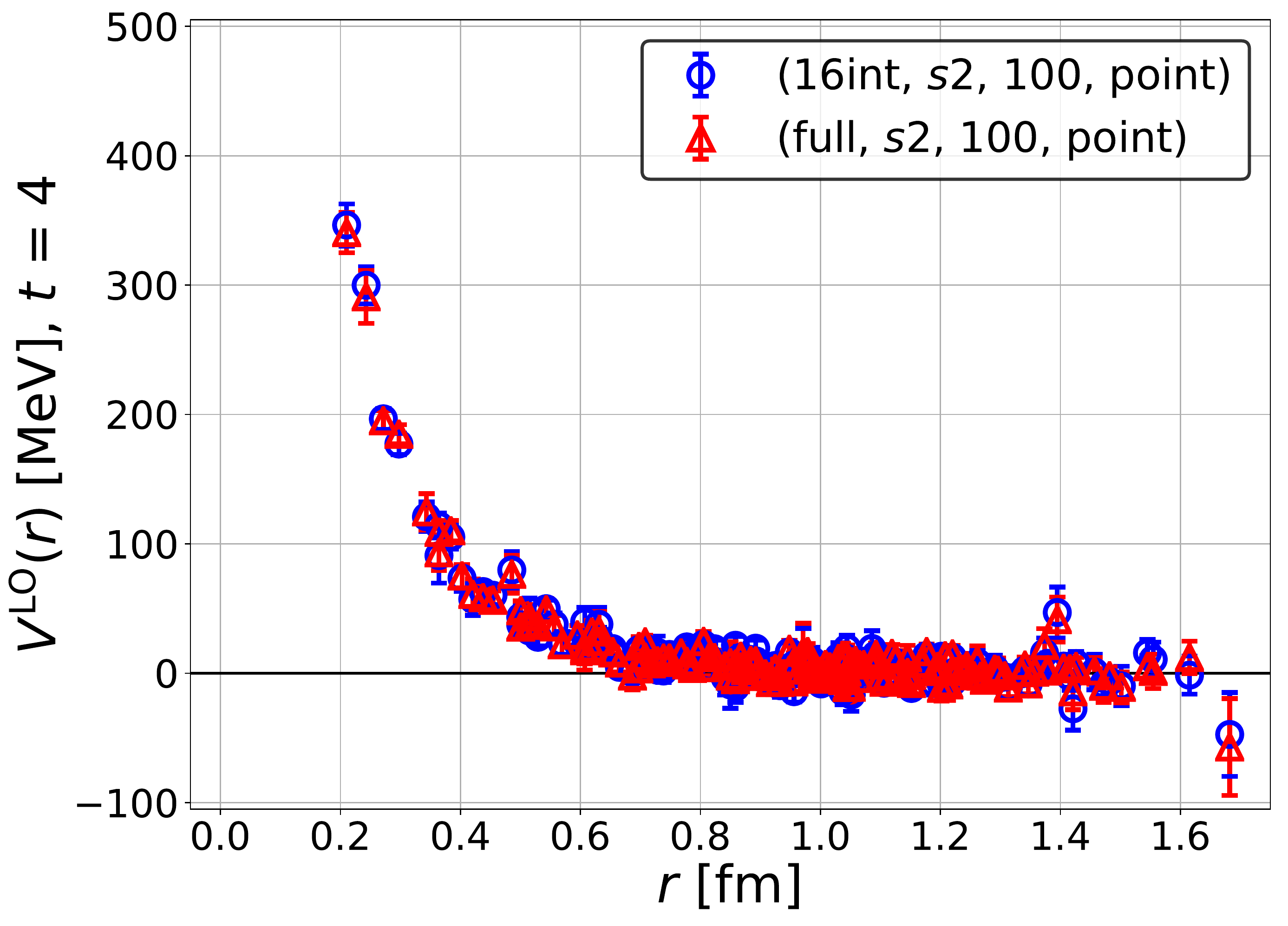}
    \end{minipage}
    & \hspace{-5mm}
    \begin{minipage}{230pt}
      \includegraphics[width=230pt,clip]{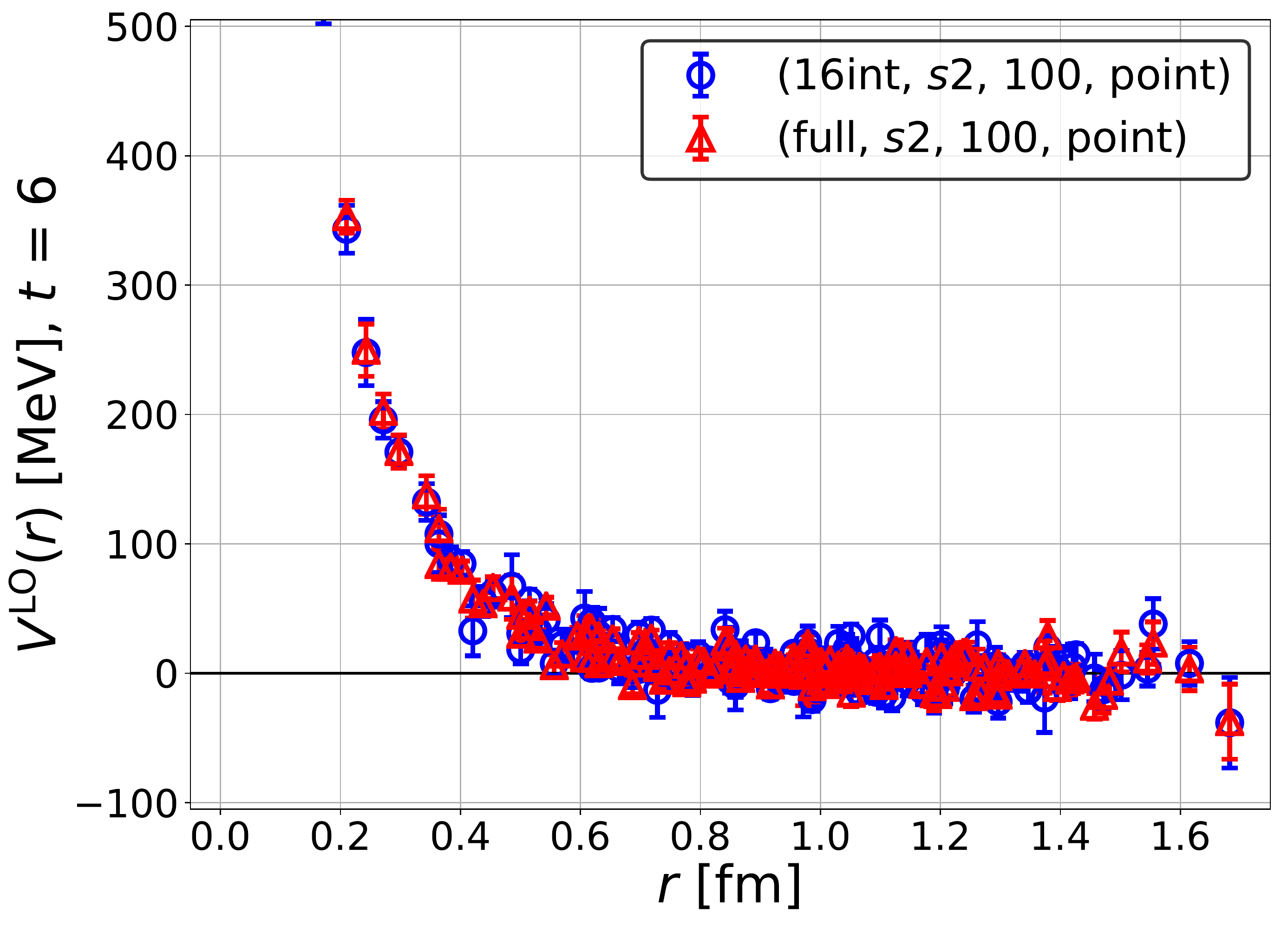}
    \end{minipage} \\
    \hspace{-7mm}
    \begin{minipage}{230pt}
      \includegraphics[width=230pt,clip]{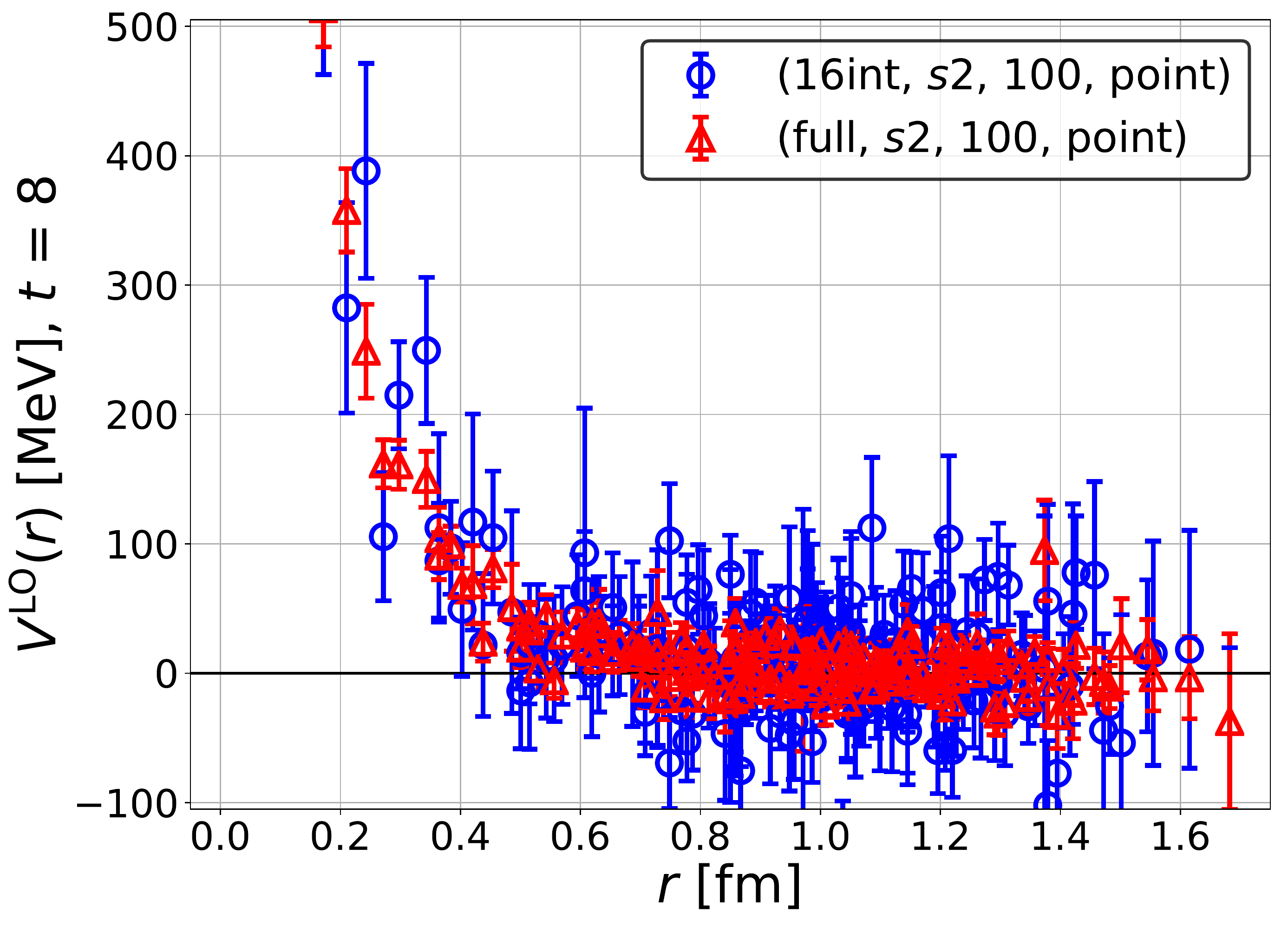}
    \end{minipage}
    & \hspace{-5mm}
    \begin{minipage}{230pt}
      \includegraphics[width=230pt,clip]{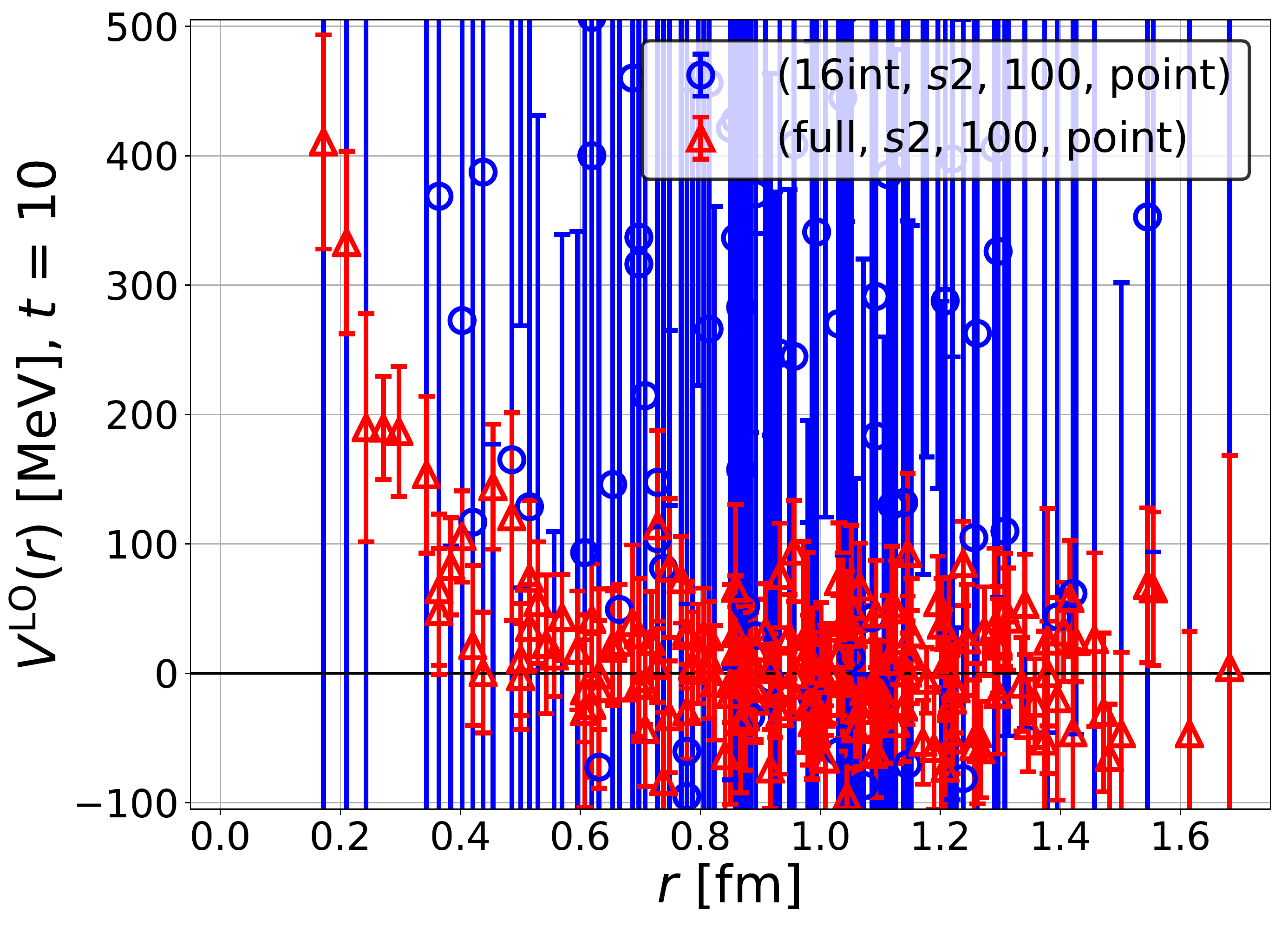}
    \end{minipage}
  \end{tabular}
  \caption{A comparison of the potentials between full (case2) (red triangles) and  16-interlace (case3) (blue circles)
    dilutions in a temporal direction at various time separations,
   $t = $ 4 (top left), 6 (top right), 8 (bottom left) and 10 (bottom right).
    }
  \label{fig:result5}
\end{figure}

In order to compensate the increased numerical cost by the introduction of the space dilution,
we investigate a possibility to reduce the cost by employing fewer dilutions in a temporal direction.
In Fig.~\ref{fig:result5},
we compare the 16-interlace time dilution (case3) with the full (32-interlace) time dilution (case2) at various $t$ ($t=4,6,8,10$)
together with the $s2$ space dilution for both cases.
We observe
that the statistical errors are comparable between the two cases at small $t$ ($t=4,6$),
while errors in the 16-interlace time dilution (case3) become much larger than those in the full time dilution (case2) at larger $t$ ($t=8,10$).

These behaviors may be qualitatively explained as follows.
If a quark propagator from $t_0$ to $t+t_0$ is estimated by the hybrid method
with the 16-interlace time dilution,
signals propagated from $t_0$ to $t+t_0$ behave as $\exp[-Et]$,
while noises contaminated from $t_0+16$ to $t+t_0$ decrease as $\exp[-E|t-16|]$.
Therefore, the signals that we need are larger than this type of noise contamination at $t < 8$.
On the other hand, the signals are largely contaminated by the  noises at $t \geq 8$, so that
the potential cannot be reliably extracted.

This observation suggests that one can reduce numerical costs
by $J$-interlace time dilutions, which reduce  $N_{\rm dil}$ to $N_{\rm dil} \times J/N_t$,
while keeping the quality of the potential, as long as the potential is calculated at $ t < J/2$.
Therefore, the most efficient way to calculate the potential would be to
combine $J$-interlace time dilution with $J$ as large as possible
together with a lattice setup that enables us to extract the potential at smaller $t$,
such as the smeared source instead of the point source.

\subsection{Effects of the smeared source}
In this subsection, we investigate
the behavior of the statistical fluctuations of the potential in the case of the smeared source.

\begin{figure}[htbp]
  \centering
  \begin{tabular}{lr}
    \hspace{-7mm}
    \begin{minipage}{230pt}
      \includegraphics[width=230pt,clip]{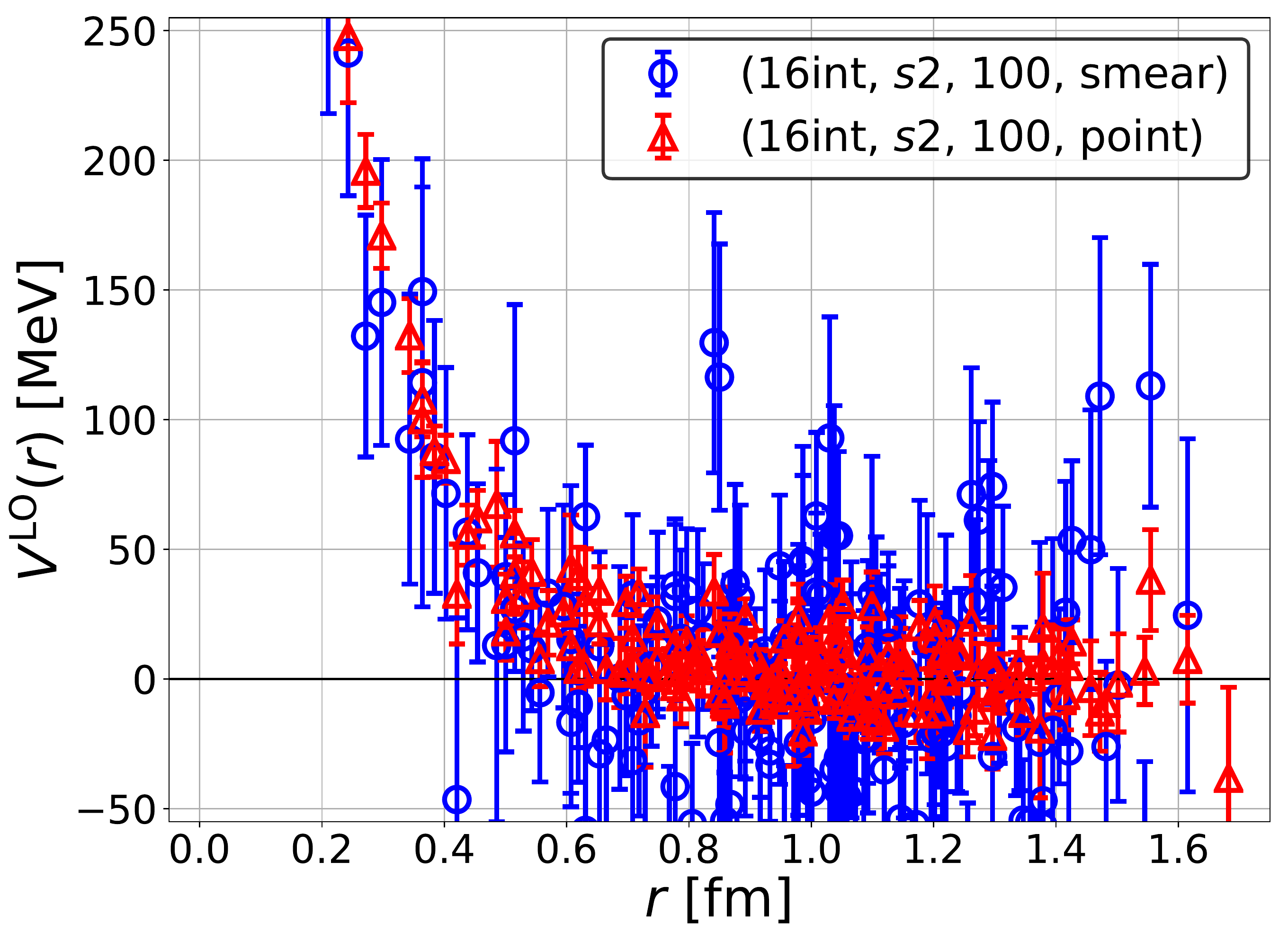}
    \end{minipage}
    & \hspace{-5mm}
    \begin{minipage}{230pt}
      \includegraphics[width=230pt,clip]{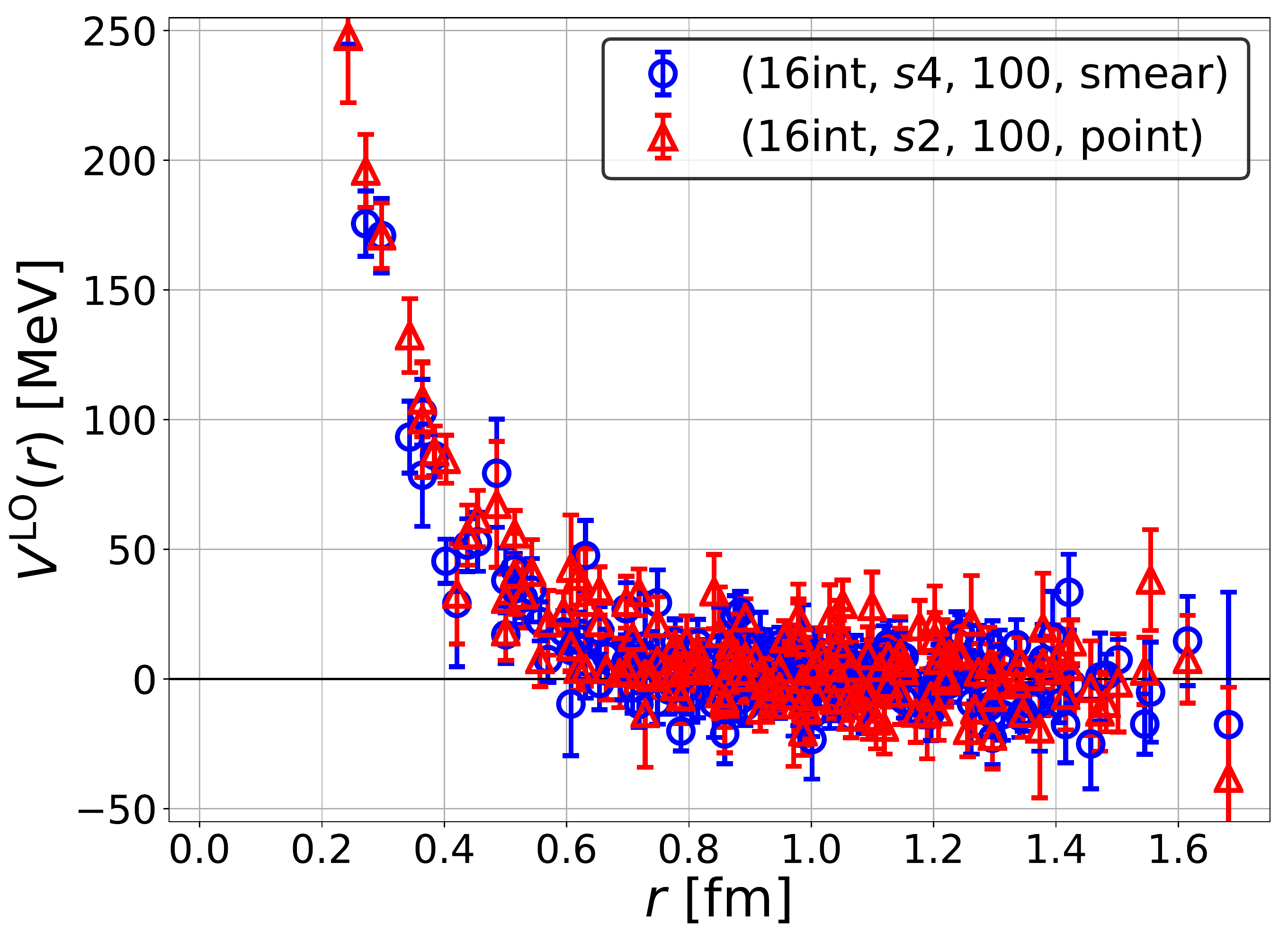}
    \end{minipage}
  \end{tabular}
  \caption{(Left) A comparison of the potential between the smeared source (case4) (blue circles) and the point source (case3) (red triangles) with 16-interlace time and $s2$ space dilution at $t=6$.
    (Right) A comparison between the smeared source with 16-interlace time and $s4$ space dilution (case5) (blue circles) and the point source with 16-interlace
    time and $s2$ space dilution (case3)(red triangles) at $t=6$.
  }
  \label{fig:result6}
\end{figure}

We first compare the potential from the smeared source (case4) with
that from the point source (case3),  keeping
16-interlace time dilution and $s2$ space dilution in both cases.
 Figure~\ref{fig:result6} (left)  shows that the source smearing makes the potential noisier.
This enhancement of noises by the smeared source may be explained by the fact
that spatial summations in the source smearing
accumulate fluctuations associated with noise vectors.
In addition, the gauge fixing may make possible cancellations among gauge-variant noises
less effective.
To reduce noise contamination due to the source smearing,
we introduce finer space dilution, $s4$.
We compare case5 (16-interlace, $s4$, 100, smear) with case3 (16-interlace, $s2$, 100, point) in  Fig.~\ref{fig:result6} (right),
which shows that a finer space dilution gives statistical errors in the potential with the smeared source comparable to those in the point source calculation.

\begin{figure}[htbp]
  \centering
  \begin{tabular}{lr}
    \hspace{-7mm}
    \begin{minipage}{230pt}
      \includegraphics[width=230pt,clip]{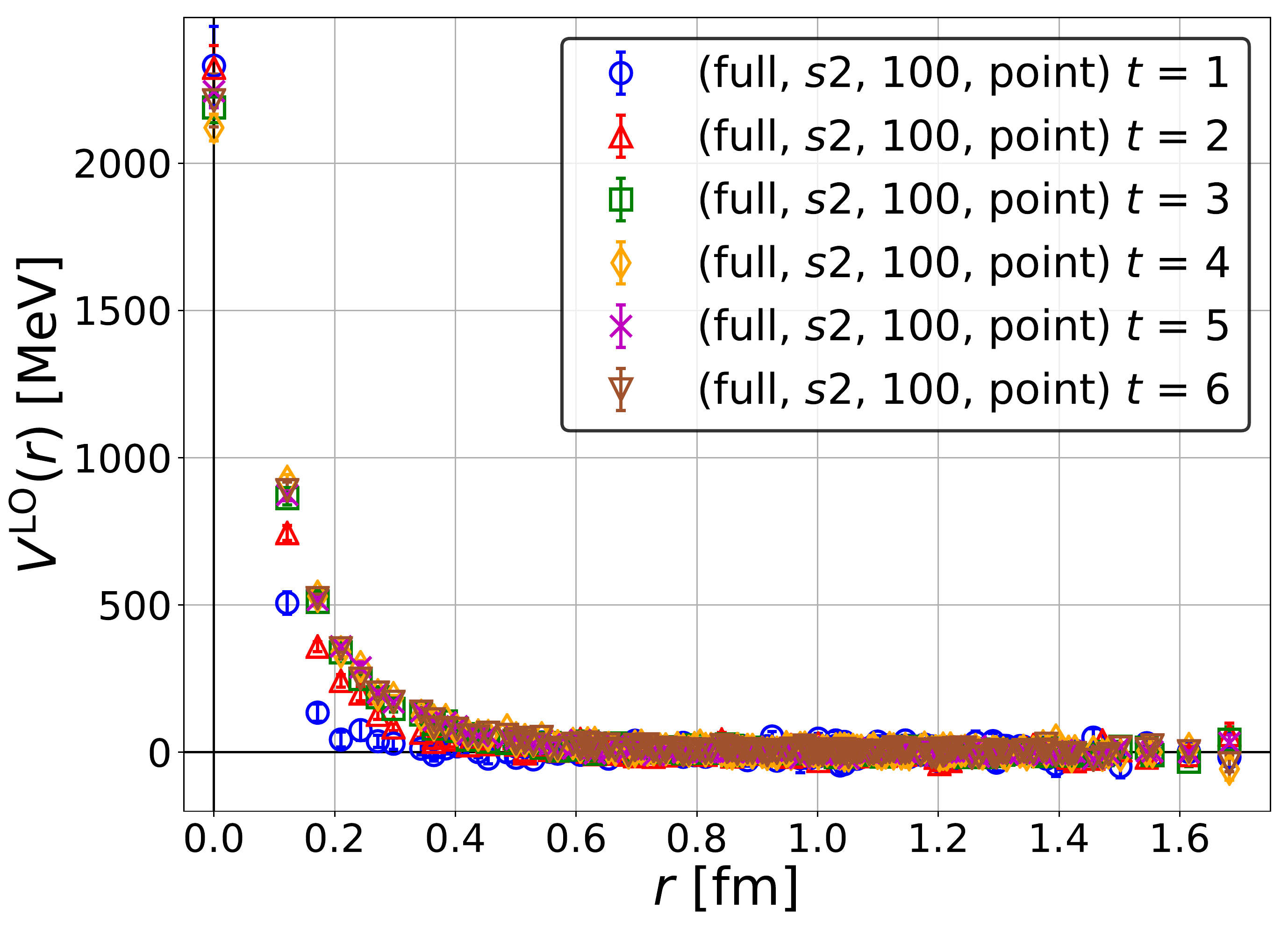}
    \end{minipage} & \hspace{-5mm}
    \begin{minipage}{230pt}
      \includegraphics[width=230pt,clip]{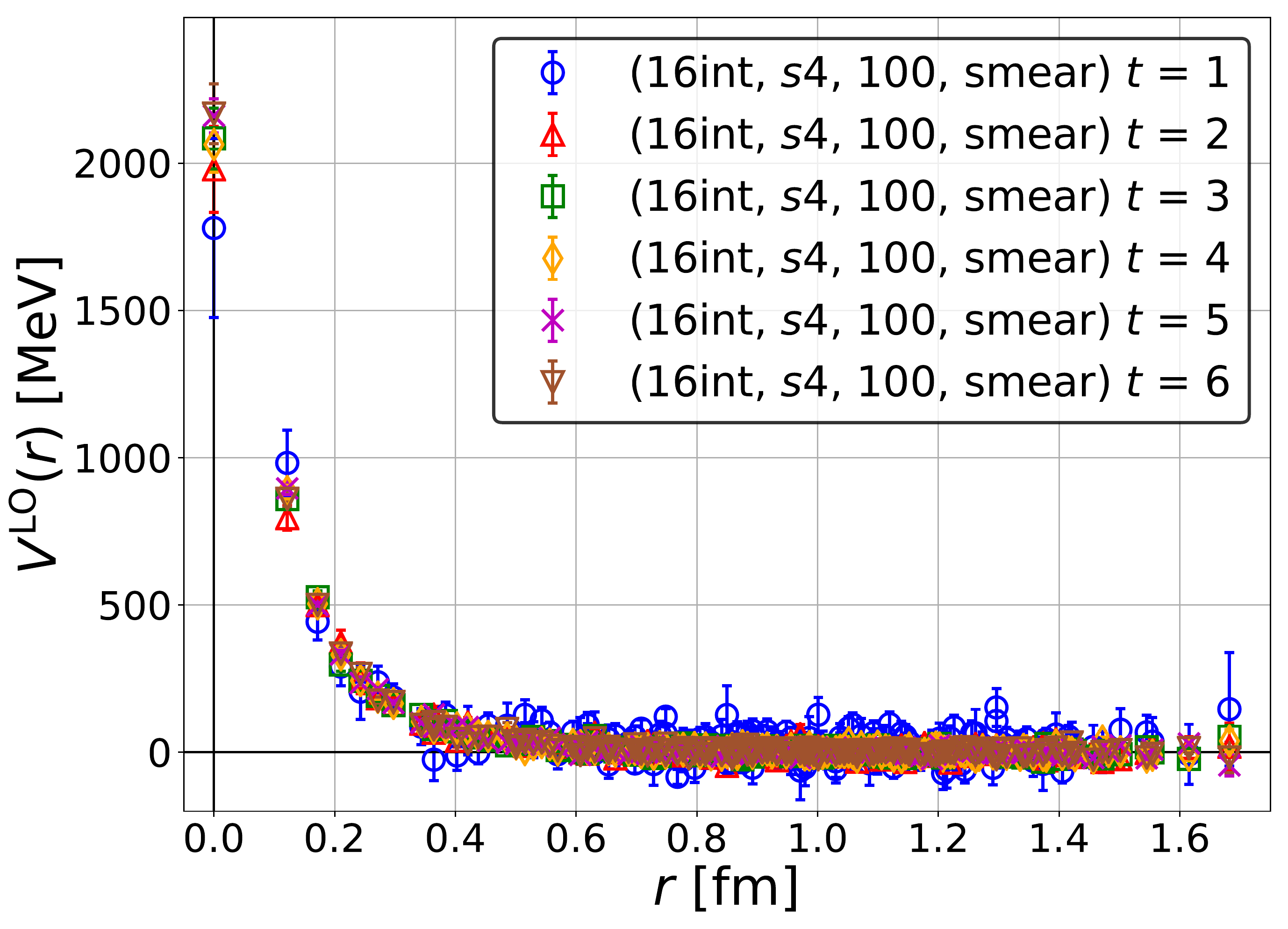}
    \end{minipage}
  \end{tabular}
  \caption{(Left) Dependence of the potential on $t$ in case2 (full, $s2$, 100, point).
  (Right) The same one in case5 (16-interlace, $s4$, 100, smear).  }
  \label{fig:result8}
\end{figure}

In Fig.~\ref{fig:result8}, we compare the time dependence of the potential with the smeared source and that with the point source. As seen in Fig.~\ref{fig:result8}, while the potential from the point source (left) has  a significant $t$ dependence at small $t$, the potential from the smeared source (right) is almost $t$-independent even at $t=2$. Therefore, the smeared source actually enables us to extract a reliable potential at an earlier time than the point source.
While the use of the smeared source may not be mandatory in the present case,
it becomes more useful when the potential from the point source shows slower convergences in time.

\subsection{Dependence on $N_{\rm eig}$}
We finally investigate how the noises of the potential depend on a number of low modes $N_{\rm eig}$.

\begin{figure}[htbp]
  \centering
  \begin{tabular}{lr}
    \hspace{-7mm}
    \begin{minipage}{230pt}
      \includegraphics[width=230pt,clip]{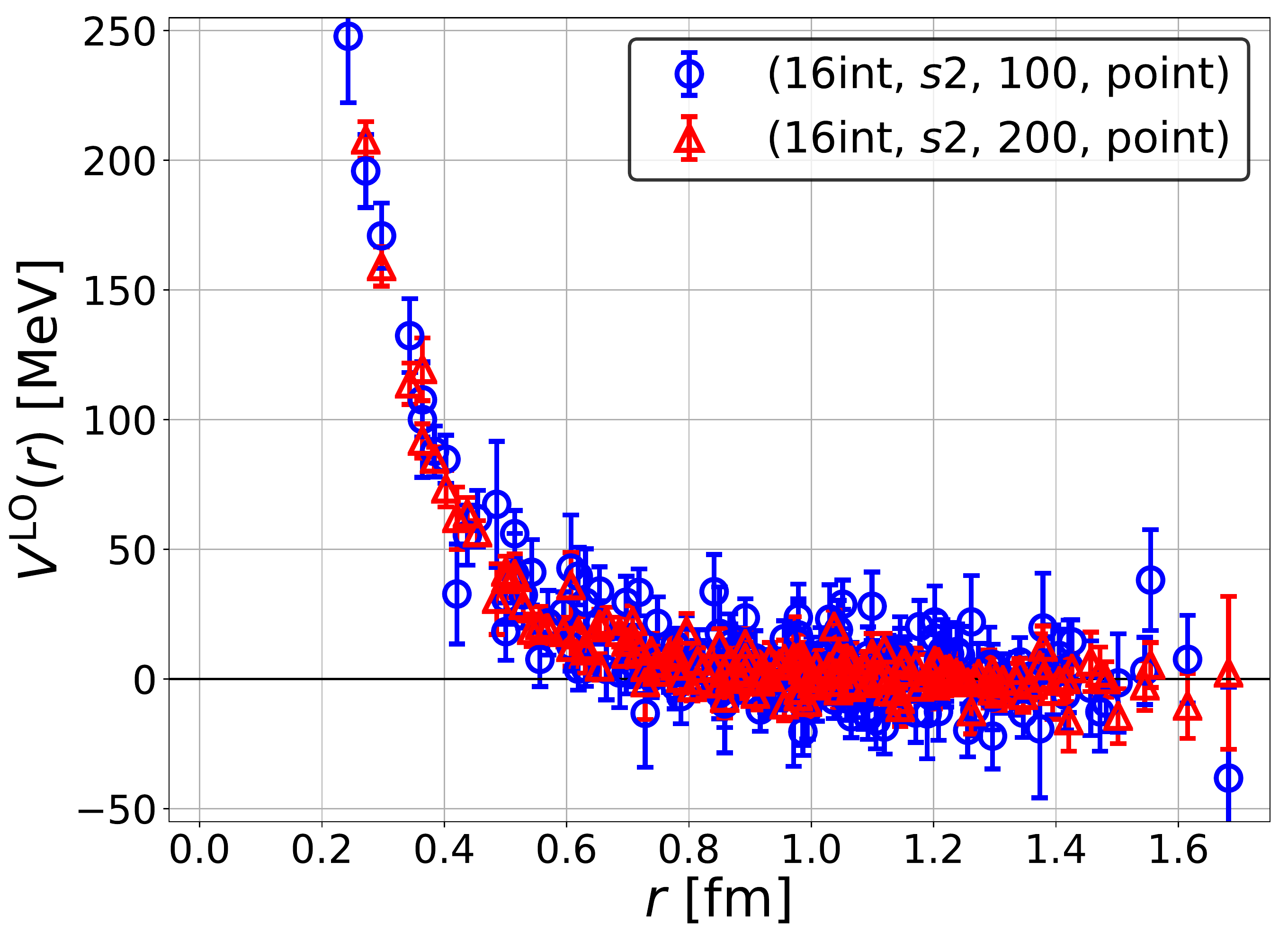}
    \end{minipage} & \hspace{-5mm}
    \begin{minipage}{230pt}
      \includegraphics[width=230pt,clip]{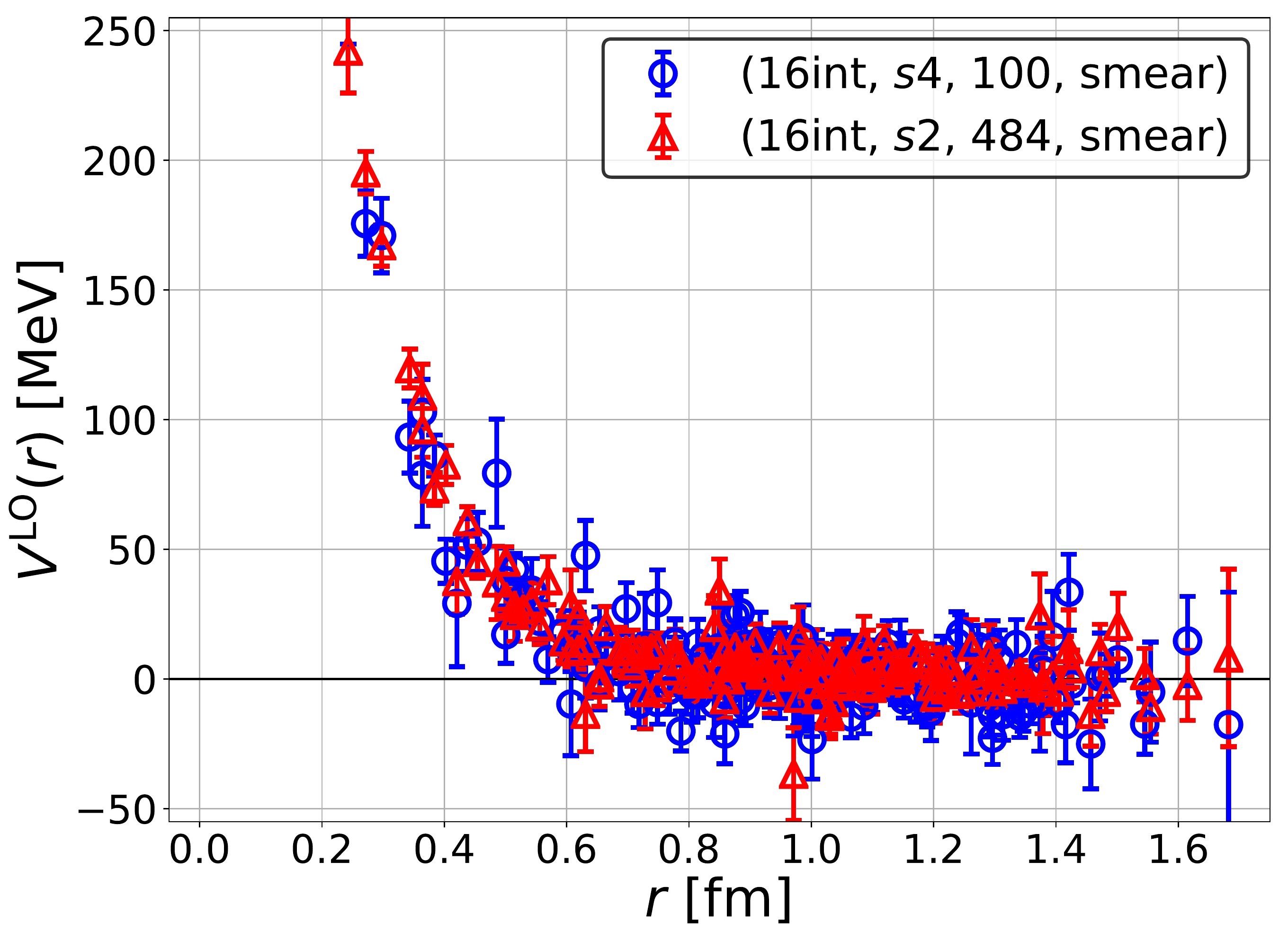}
    \end{minipage}
  \end{tabular}
  \caption{(Left) A comparison of the potential between case3, $N_{\rm eig} = 100$ (blue) and case6,  $N_{\rm eig} = $ 200 (red)
  with the same (16-interlace, $s2$, point).
  (Right) A comparison of the potential between case5, (16-interlace, $s4$, smear) with $N_{\rm eig} = 100$
  (blue),  and case7, (16-interlace, $s2$, smear) with $N_{\rm eig} = 484$ (red).
  Both correspond to the same $N_{\rm hl} = N_{\rm eig} + N_{\rm dil} = 868$. }
  \label{fig:result9}
\end{figure}

Naively, we expect that statistical errors become smaller for larger $N_{\rm eig}$, since the relative segment of the propagator estimated exactly, $D_0^{-1}$, becomes larger.
In order to confirm this point explicitly,
we compare potentials at $t=6$ between $N_{\rm eig} = 100$ (case3) and $200$ (case6) with (16-interlace, $s2$, point)
in  Fig.~\ref{fig:result9} (left),
which shows that the potential with $N_{\rm eig} = 200$ (red) has smaller noise contamination than the one
with $N_{\rm eig} = 100$ (blue). Therefore, we can confirm that our expectation is indeed the case.

Next, in order to see which is more important for noise reductions, finer space dilution or larger $N_{\rm eig}$, we compare  (16-interlace, $s2$, $N_{\rm eig} = 484$, smear) (case7) with
(16-interlace, $s4$, $N_{\rm eig} = 100$, smear) (case5),
keeping $N_{\rm hl} = N_{\rm eig} + N_{\rm dil} = 868$ the same for both cases. Note that, in our setup, $N_{\rm hl}$ is effectively a good measure for numerical costs because the most time-consuming part of our calculations is the contraction part, whose numerical costs are controlled by $N_{\rm hl}$.
Figure~\ref{fig:result9} (right) indicates that a
larger $N_{\rm eig}$ (case7, red) is a little better for the potential to have smaller noise contamination than a
smaller $N_{\rm eig}$ (case5, blue).

From these studies, taking larger $N_{\rm eig}$ is slightly advantageous over finer space dilutions in our case. However, this relative efficiency could depend on the actual value of $N_{\rm eig}$ and the lattice setup such as the size of the lattice volume.
In particular,
we cannot freely make $N_{\rm eig}$ larger and larger,
since the numerical costs for the calculation of eigenmodes become
non-negligible at some point.
Therefore, increasing $N_{\rm eig}$ as long as the numerical cost for the eigenmodes remains subdominant
will be the first guiding principle
before performing the detailed optimization on $N_{\rm eig}$.

\subsection{Lessons in this section}
\label{sec:lesson}
In order to extract the HAL QCD potential with the hybrid method for all-to-all propagators,
lessons learned from the investigations in this section are summarized as follows.\vspace{5mm}\\
(1) A finer space dilution should be used to reduce noise contamination to the potential, in addition to full color and spinor dilutions.  \\
(2) The smeared source accumulates noise contamination; thus additional dilutions in spatial directions are mandatory. We should take the smeared source to extract the potential at the smallest possible $t$ if potentials with the point source become reliable only at larger $t$. \\
(3) A $J$-interlace time dilution can be used for the potential to be extracted at $t < J/2$,
to reduce $N_{\rm dil}$  by a factor $N_t/J$. \\
(4) It is better to increase $N_{\rm eig}$
until the costs for the calculation of eigenmodes become significant.
 The total number of $N_{\rm hl}$ becomes $N_{\rm hl} = 12\times J\times  2^{s/2} + N_{\rm eig}$,
 where $s$ corresponds to a level of the space dilution, $s=0$ (no dilution), $s=2$ ($s2$),
 $s=4$ ($s4$).

\section{Comparison with the result without all-to-all propagators}
In this section,
we compare  the $I=2$ $\pi\pi$ potential and the corresponding scattering phase shifts obtained in the hybrid method for all-to-all propagators
with those in the standard method without  all-to-all propagators.

\begin{figure}[htbp]
  \centering
  \begin{tabular}{lr}
    \hspace{-7mm}
    \begin{minipage}{230pt}
      \includegraphics[width=230pt,clip]{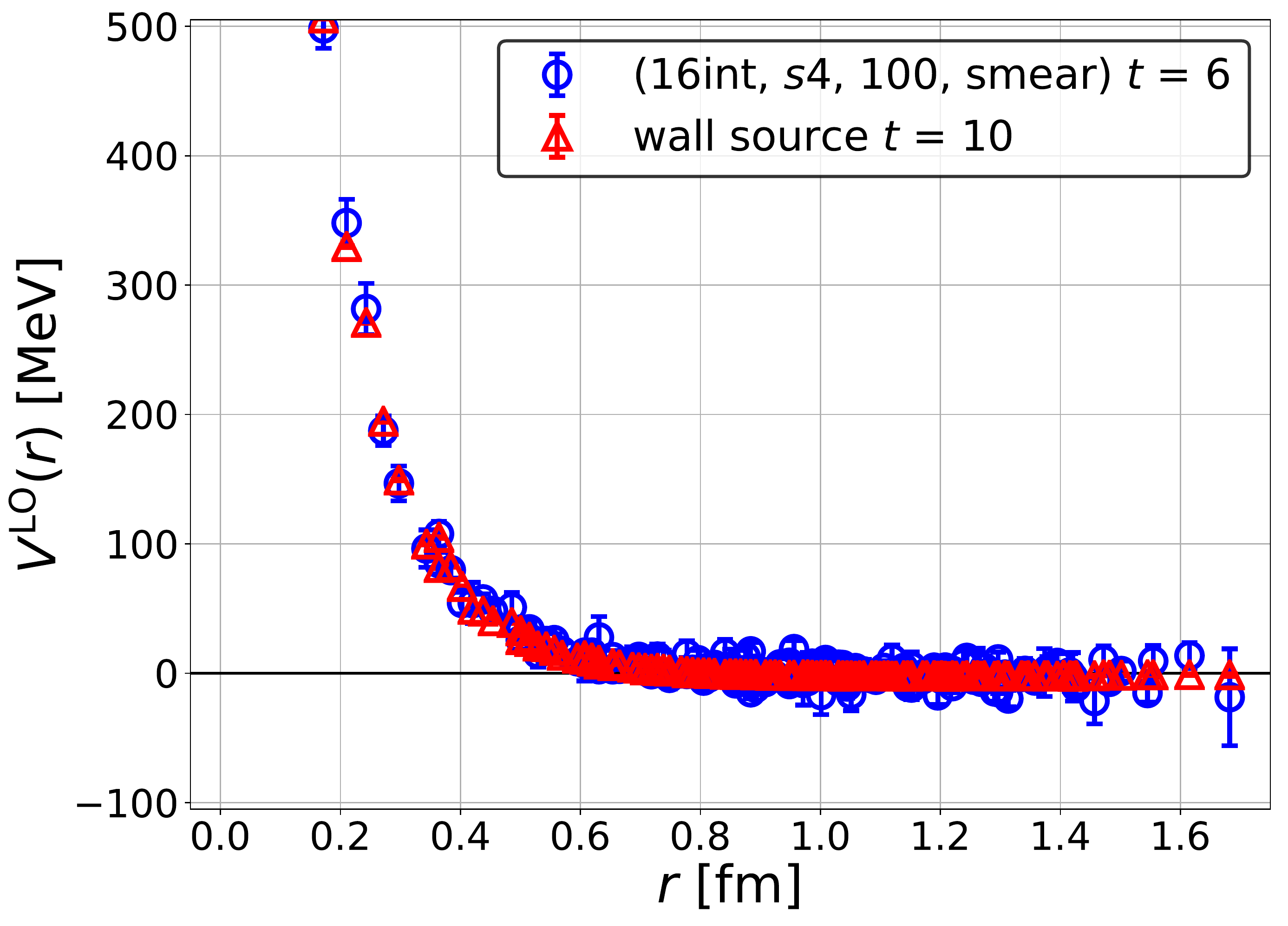}
    \end{minipage} & \hspace{-5mm}
    \begin{minipage}{230pt}
      \includegraphics[width=230pt,clip]{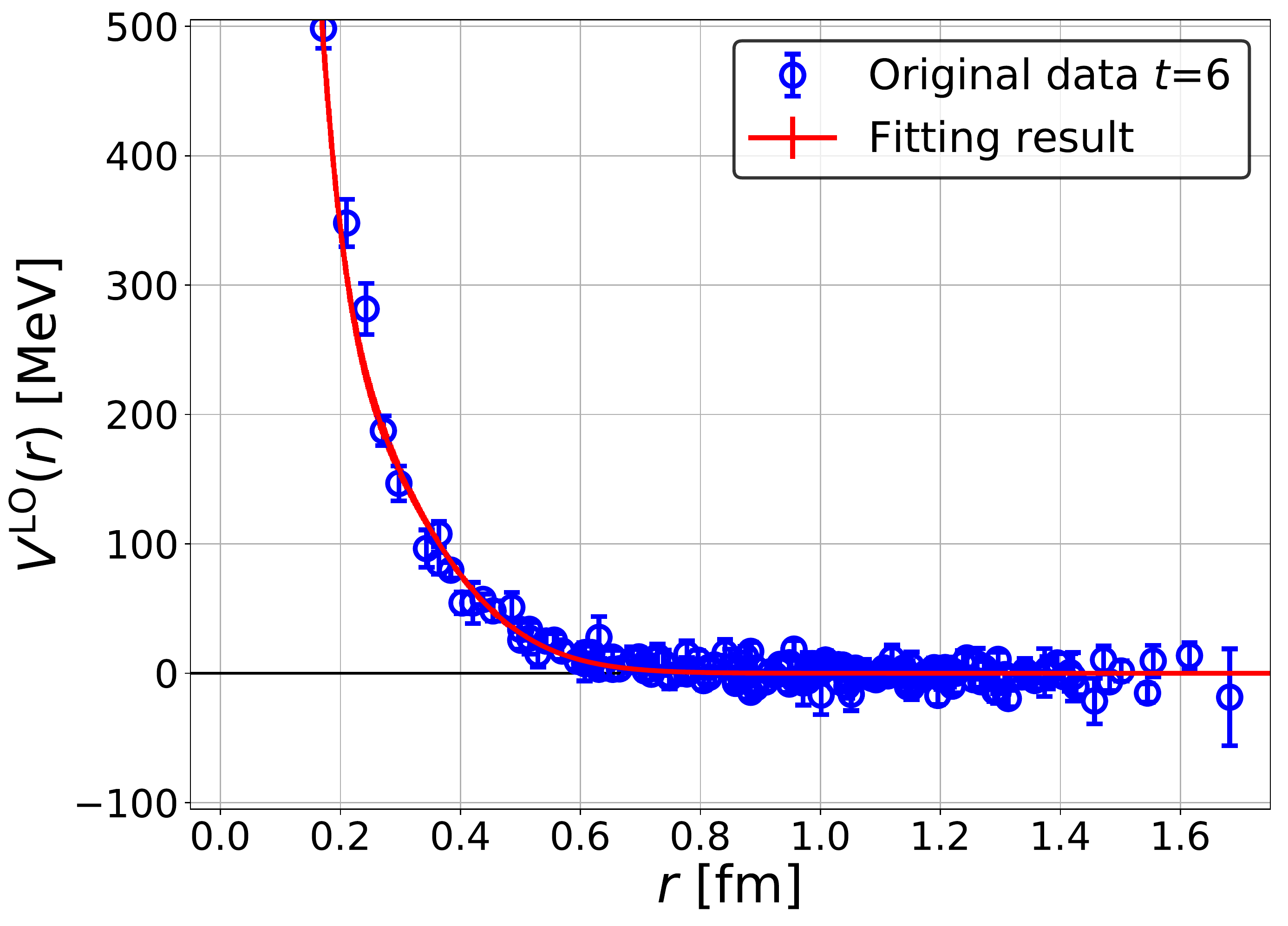}
    \end{minipage}
  \end{tabular}
  \caption{(Left) A comparison of the $I=2$ $\pi\pi$ potentials,
    one from the hybrid method (blue circles) and the other from the wall quark source (red triangles).
    (Right)
  The result of the fit (red line) of the potential from the hybrid method (blue circles).}
  \label{fig:result14}
\end{figure}

Figure~\ref{fig:result14} (left) compares the LO potential for the $I=2$ $\pi\pi$ system
obtained by the hybrid method (16-interace, $s4$, $N_{\rm eig}=100$, smear) with $N_{\rm conf}=60$ (case5a)  at $t=6$ with the one with the conventional setup in the HAL QCD method
(32 wall quark sources/conf with $N_{\rm conf}=700$) at $t=10$.
Both results agree with each other within statistical errors.

Then the potential is fitted by
\begin{equation}
  V(r) = a_0 e^{-(r/a_1)^2} + a_2 e^{-(r/a_3)^2},
\end{equation}
from which the $I=2$ $\pi\pi$ scattering phase shifts  are extracted.
Figure~\ref{fig:result14} (right) shows the fit line, and
 Table~\ref{tab:result5} summarizes the fit parameters.
\begin{table}[btp]
  \caption{Fit parameters for the potential, $a_i$, and
    the corresponding $\chi^2/{\rm d.o.f.}$.
  }
  \vspace{2mm}
  \label{tab:result5}
  \centering
  \begin{tabular}{cccc|c}
  \hline
    $a_0$ [MeV] & $a_1$ [fm] & $a_2$ [MeV] & $a_3$ [fm] & $\chi^2/{\rm d.o.f.}$ \\ \hline
    2050(30) & 0.113(0.002) & 380(26) & 0.316(0.008) & 1.27\\
    \hline
  \end{tabular}
\end{table}

\begin{figure}[htbp]
  \centering
  \begin{tabular}{lr}
    \hspace{-7mm}
    \begin{minipage}{230pt}
      \includegraphics[width=230pt,clip]{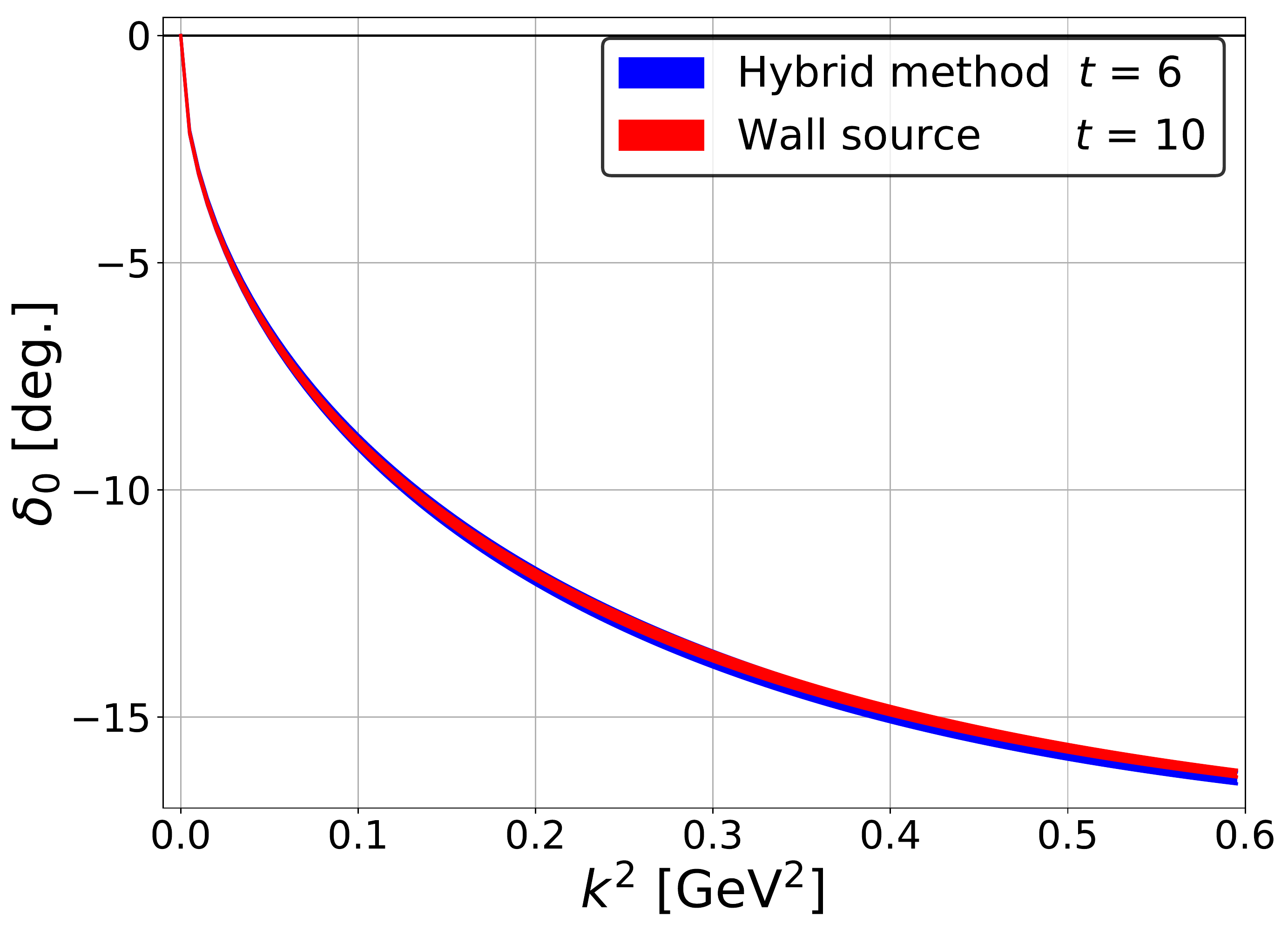}
    \end{minipage} & \hspace{-5mm}
    \begin{minipage}{230pt}
      \includegraphics[width=230pt,clip]{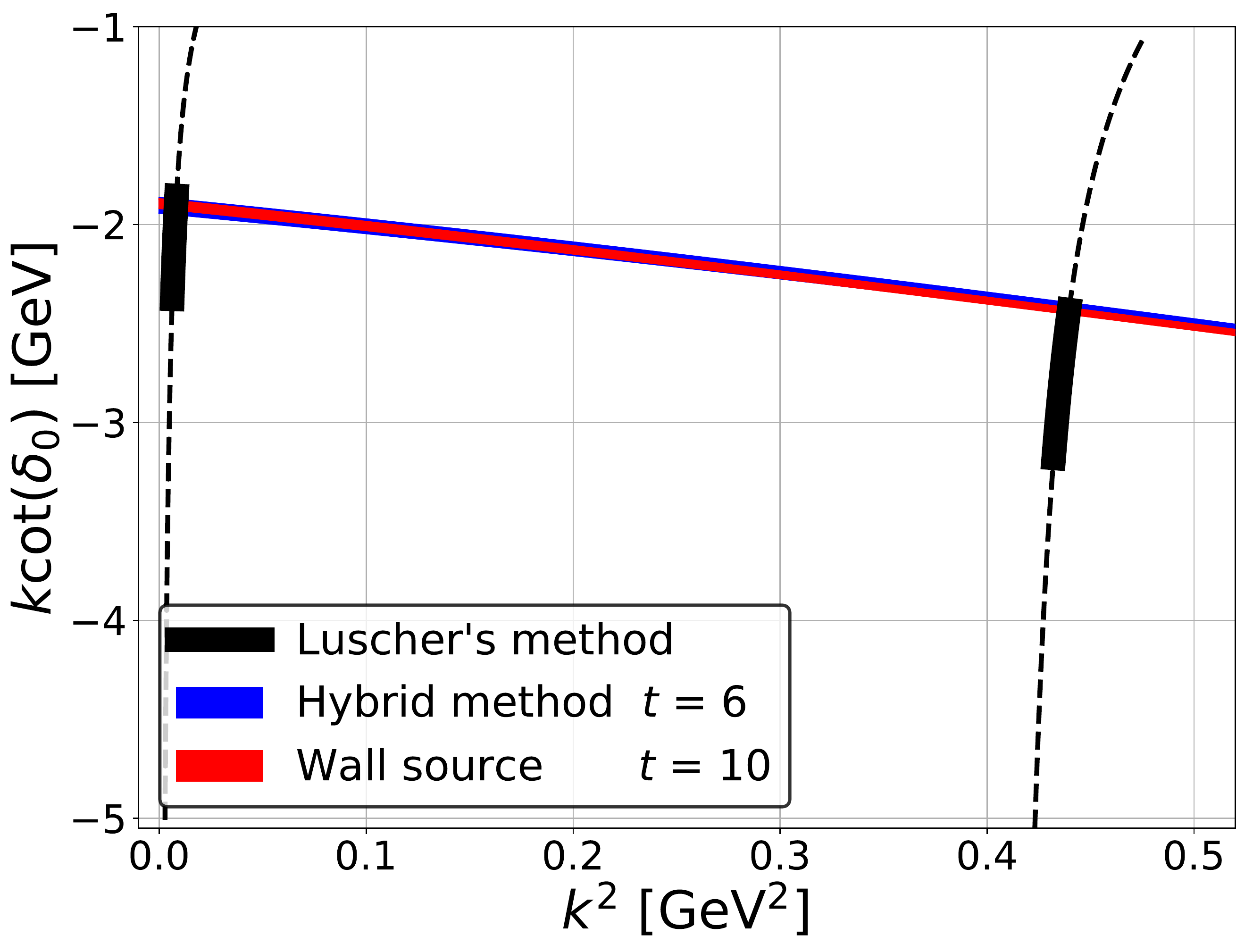}
    \end{minipage}
  \end{tabular}
  \caption{(Left) Scattering phase shifts $\delta_0 (k)$ as a function of $k^2$.
    (Right) $k\cot\delta_0(k)$ as a function of $k^2$.
    Blue (red) bands correspond to the results from the HAL QCD method
      with the hybrid method (with the wall quark source).
    Black bands in the right figure correspond to
   the results from L\"uscher's method~\cite{hal_kawaisan}.}
  \label{fig:result15}
\end{figure}

In Fig.~\ref{fig:result15},
we present the $I=2$ $\pi\pi$ scattering phase shifts $\delta_0(k)$ (left) and $k\cot\delta_0(k)$
(right) as a function of $k^2$,
together
with
results from the HAL QCD method with the wall quark source and those from
 L\"uscher's finite volume method \cite{hal_kawaisan}.
As expected from the agreement of the potential, we confirm that the results by the hybrid method agree with the ones without all-to-all propagators (wall quark source)  and the results from L\"uscher's method.

\section{Conclusion}
In this paper, we employ the hybrid method of all-to-all propagators\cite{hybrid} for the HAL QCD method and
study the interaction of the $I=2$ $\pi\pi$ system at $m_{\pi} \approx 870$ MeV.
Even though the hybrid method brings extra statistical fluctuations for the results,
we obtain a reasonably accurate potential by increasing dilution levels, which gives $I=2$ $\pi\pi$ scattering phase shifts consistent with the result using the conventional method.
Our findings for appropriate choices of parameters in the hybrid method to calculate the potentials
are summarized in Sect.~\ref{sec:lesson}.

As the hybrid method works in the HAL QCD method,
we will calculate the $I=1$ $\pi\pi$ potential using all-to-all propagators.
It is interesting to see whether the $\rho$ resonance is correctly reproduced by the HAL QCD potential. A preparatory study has already been made, and the results will be published in the near future.
Future applications include unconventional hadrons such as
$\sigma$/$f_0(500)$, exotic $X, Y, Z$ states\cite{Lebed:2016hpi} and pentaquark states $P_c$\cite{Aaij:2015tga,Aaij:2019vzc},
whose natures are not yet understood.
The importance of first-principles lattice QCD study is increasing more than ever,
and it is expected that our future studies with a combination of the HAL QCD method
and all-to-all propagators will shed light on these unconventional states.

\section*{Acknowledgements}
The main part of our calculation code is based on the Bridge++ code\cite{bridgepp1,bridgepp2}.
We thank the JLQCD and CP-PACS Collaborations for providing their 2+1 configurations\cite{jlqcd,cppacsjlqcd}.
All the numerical calculations were performed on the Cray XC40 at the Yukawa Institute for Theoretical Physics (YITP), Kyoto University.
This work is supported in part by JSPS Grants-in-Aid for Scientific Research, No. JP19K03879, JP18H05236, JP18H05407, JP16H03978, JP15K17667, by a priority issue (Elucidation of the fundamental laws and evolution of the universe) to be tackled by using the Post ``K" Computer,
and by the Joint Institute for Computational Fundamental Science (JICFuS).
The authors thank the members of the HAL QCD Collaboration for fruitful discussions.


\end{document}